\documentclass[a4paper,amsmath,amssymb,aps,prl,twocolumn,longbibliography,10pt,
floatfix,
english,superscriptaddress]{revtex4-2}

\usepackage{amssymb}
\usepackage{graphicx}
\usepackage{dcolumn}
\usepackage{bm}
\usepackage{amsmath}
\usepackage{subfigure}
\usepackage{float}
\usepackage{color}
\usepackage[colorlinks=true,citecolor=blue]{hyperref}
\usepackage{multirow}

\usepackage{xr}

\begin{document}

\preprint{APS/123-QED}

\title{Thermal resonance-enhanced transparency in room temperature Rydberg gases} 
	
\author{Jinlian Hu}
\thanks{These authors contributed equally to this work.}
\affiliation{State Key Laboratory of Quantum Optics Technologies and Devices, Institute of Laser Spectroscopy, Shanxi University, Taiyuan 030006, China}
 
\author{Yuechun Jiao}
\thanks{These authors contributed equally to this work.}
\email{ycjiao@sxu.edu.cn}
\affiliation{State Key Laboratory of Quantum Optics Technologies and Devices, Institute of Laser Spectroscopy, Shanxi University, Taiyuan 030006, China}
\affiliation{Collaborative Innovation Center of Extreme Optics, Shanxi University, Taiyuan 030006, China}
 
\author{Yuwen Yin}
\affiliation{State Key Laboratory of Quantum Optics Technologies and Devices, Institute of Laser Spectroscopy, Shanxi University, Taiyuan 030006, China}
 
\author{Cheng Lu}
\affiliation{State Key Laboratory of Quantum Optics Technologies and Devices, Institute of Laser Spectroscopy, Shanxi University, Taiyuan 030006, China}

\author{Jingxu Bai}
\affiliation{State Key Laboratory of Quantum Optics Technologies and Devices, Institute of Laser Spectroscopy, Shanxi University, Taiyuan 030006, China}
\affiliation{Collaborative Innovation Center of Extreme Optics, Shanxi University, Taiyuan 030006, China}

\author{Suotang Jia}
\affiliation{State Key Laboratory of Quantum Optics Technologies and Devices, Institute of Laser Spectroscopy, Shanxi University, Taiyuan 030006, China}
\affiliation{Collaborative Innovation Center of Extreme Optics, Shanxi University, Taiyuan 030006, China}

\author{Weibin Li}
\email{weibin.li@nottingham.ac.uk}
\affiliation{School of Physics and Astronomy and Centre for the Mathematics and Theoretical Physics of Quantum Non-equilibrium Systems, University of Nottingham, Nottingham NG7 2RD, United Kingdom}

\author{Zhengyang Bai}
\email{zhybai@nju.edu.cn}
\affiliation{National Laboratory of Solid State Microstructures and School of Physics,
Collaborative Innovation Center of Advanced Microstructures, Nanjing University, Nanjing 210093, China} 
\affiliation{Hefei National Laboratory, Hefei 230088, China}

\author{Jianming Zhao}
\email{zhaojm@sxu.edu.cn}
\affiliation{State Key Laboratory of Quantum Optics Technologies and Devices, Institute of Laser Spectroscopy, Shanxi University, Taiyuan 030006, China}
\affiliation{Collaborative Innovation Center of Extreme Optics, Shanxi University, Taiyuan 030006, China}
	
\date{\today}

\begin{abstract}
We report the enhanced optical transmission in the coherent, \textit{off-resonant} excitation of Rydberg atom gases at room temperature via a two-photon process. Here thermal resonance-enhanced transparency (TRET) is induced when the detuning of the two lasers is adjusted to compensate the atomic thermal-motion-induced energy shifts, i.e. single and two-photon Doppler shifts. We show that the atomic velocity is mapped into the transmission of the probe fields, which can be altered by independently and selectively exciting different velocity groups through sweeping the detuning. The maximal transmission in TRET is about 8 times higher than that under the electromagnetically induced transparency (EIT). Utilizing the TRET effect, we
enhance the sensitivity of a Rydberg microwave receiver to be 28.7~nVcm$^{-1}$Hz$^{-1/2}$, ultimately reaching a factor of 2.1 of the EIT case. 
When atoms of separate velocity groups are excited simultaneously by 
multiple sets of detuned lasers, the receiver sensitivity further increases, which is linearly proportional to the number of the velocity groups. Our study paves a way to exploit light-matter interaction via the TRET, and contributes to current efforts in developing quantum sensing, primary gas thermometry, and wireless communication with room-temperature atomic gases. 
\end{abstract}


\maketitle
\textit{Introduction} -- 
Enhancement of atom-light interaction is one of the most challenging endeavors for a variety of research fields~\cite{RevModPhys.77.633}, and has driven significant progress in quantum information, simulation, and metrology~\cite{saffman2010,nohQuantumSimulationsManybody2016,adams2019}. One strategy to achieve strong coupling regime is to confine atoms to high-finesse optical cavities to boost the interaction probability~\cite{aoki2006,lepert2011}. In free space, one utilizes the collective coupling of photons in ensembles consisting of many atoms~\cite{bettles2016,ferioli2021}. The coupling efficiency is maximized under the resonant or near-resonant excitation~\cite{boller1991,fleischhauer2005a,chu2025,liu2021,allen2012,abi2010}. Among many protocols,  electromagnetically induced transparency (EIT)~\cite{boller1991,fleischhauer2005a} has emerged as a key technique for achieving strong light-matter interaction for studying enhancement of optical transmission~\cite{fleischhauer2005a}, slow light~\cite{chu2025}, and even storing photons~\cite{phillips2001} in cold atom gases.

\begin{figure*}[htbp]
    \centering
\includegraphics[width=1\textwidth]{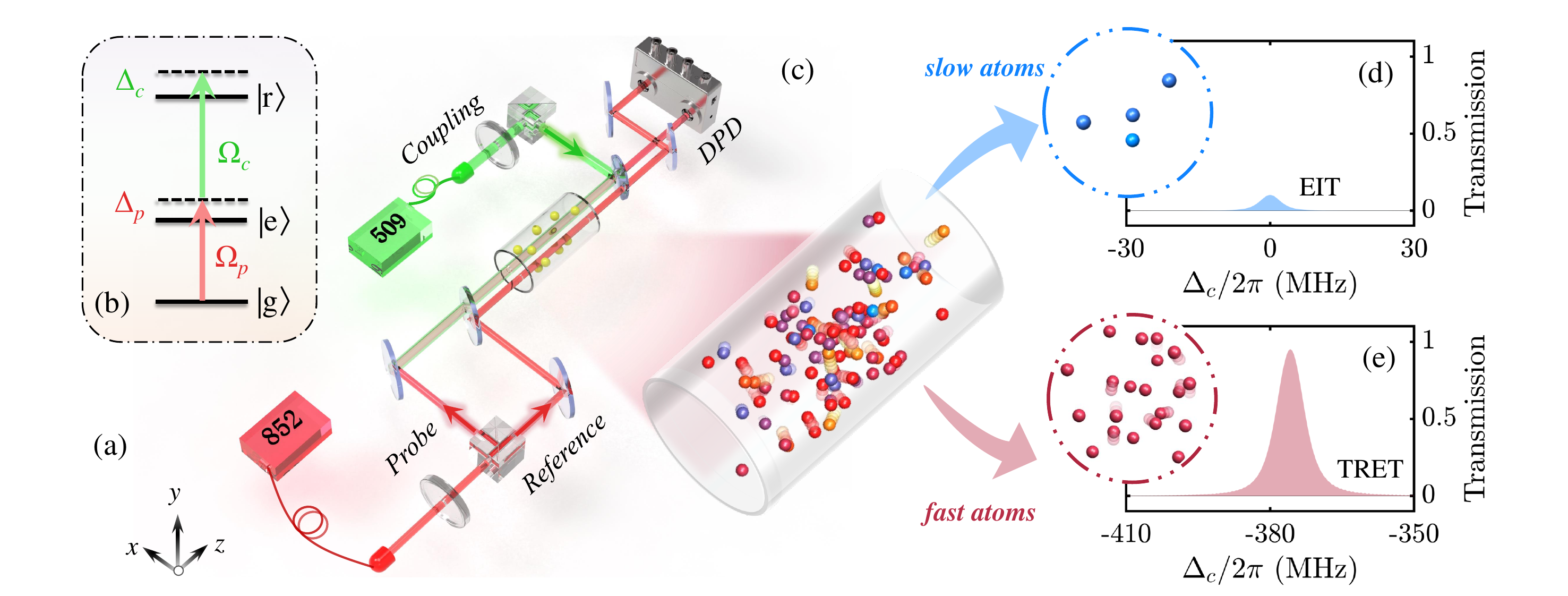}
    \caption{TRET in thermal Rydberg gases. (a) Sketch of the experimental setup and (b) relevant energy-level diagram. An 852 nm laser is split into two identical beams, labeled as the reference beam and the probe beam. The 509 nm coupling laser counter-propagates through the vapor cell, and overlaps with the probe laser but not the reference beam. The probe laser (Rabi frequency $\Omega_p$) and the coupling laser ($\Omega_c$) drive the ground atoms $|g\rangle$ to the Rydberg state $|r\rangle$ via an intermediate state $|e\rangle$ with detuning $\Delta_p$. (c) The thermal motion of atoms in the cell follows the Maxwell–Boltzmann distribution $N(v)\sim v^2f(v)$. (d) and (e) Transmission when $\Delta_p/2\pi =0$~MHz and $\Delta_p=225$~MHz. At these detunings the lasers selectively excite velocity groups such that the peak transmission appears at velocity $v_c \simeq 0$ m/s and $v_c \approx  v_T = 193$ m/s, correspondingly.
    }
    \label{figure1}
\end{figure*}

At room temperature atom-light interaction is critically important for real-world applications where a cryogenic environment is avoided.
However random thermal motions of atoms inevitably weaken the desired quantum effects~\cite{gea-banacloche1995a, wu2008} due to the Doppler and collisional shifts.
To mitigate the thermal effect, one approach is to excite hot atoms with short and strong lasers~\cite{urvoy2015, zhang2016a, baluktsian2013, bai2020}, where thermal motions are relatively weak on the nanosecond (or shorter) time scale.
On the other hand, the Doppler effect is nature’s silent symphony of motion, that links the thermal energy to an optical frequency via laser spectroscopy, achieving such as non-reciprocal optical devices~\cite{zhang2018a, liang2020, dong2021} and atomic frequency combs~\cite{aumiler2005, afzelius2009, main2021}. In the off-resonant regime, the laser frequency mismatches the transition. The extra photon energy, on the other hand, can compensate the Doppler effect, i.e. the thermal motion induced energy shift, leading to typically narrow resonant excitation. How to exploit this thermal resonance and find its quantum technological applications are interesting and have not been fully explored.

In this work, we investigate coherent, off-resonant Rydberg excitation of thermal cesium atoms in a vapor cell. We consider a Rydberg EIT setting, where a probe and coupling laser couple ground $|g\rangle$, intermediate $|e\rangle$ and Rydberg state $|r\rangle$, as depicted in Figs.~\ref{figure1}(a) and (b).
At room temperature,
selective velocity groups of atoms are excited when the detuning of the probe and coupling lasers compensate the single and two photon Doppler shifts. Under this condition, the probe transmission reaches its peak values around a narrow region centered at the selected velocity,
leading to a {\it thermal resonance-enhanced transparency} (TRET). We show that the maximal transmission in the TRET is more than 8 times that of the EIT regime with resonant laser excitation. The enhanced transmission and narrow TRET allow quantum technological applications in sensing microwave (MW) electric fields. This is demonstrated experimentally with a MW receiver, where the sensitivity is improved to be 28.7~nVcm$^{-1}$Hz$^{-1/2}$, a factor of 2.1 better than in the resonant EIT.
By exciting multiple velocity groups of atoms simultaneously, the sensitivity can be further improved, which is approximately proportional to the velocity groups number. The TRET thus can enhance the performance of Rydberg atom field sensing~\cite{sedlacek2012,Jing2020,simons2021,schlossberger2024,elgee2024,robinson2021determining,holloway2014subwave,meyer2021} and could boost Rydberg quantum technologies in, for example, communications~\cite{cox2018Quantum, anderson2021, simons2019b, holloway2019}.  

\textit {Model} --
In our Rydberg-EIT setting two counterpropagating probe and coupling lasers  (wave vector $\mathbf{k}_p$ and $\mathbf{k}_c$ along the $z$ axis) propagating in the room-temperature  
Cs gas (density ${\cal N}$),  coupling the ground state $|g\rangle$ to the Rydberg state
$|r\rangle$ via an intermediate state $|e\rangle$, as depicted Figs.~\ref{figure1}(a) and (b). 
Under the rotating-wave approximation, this yields the  Hamiltonian of the system ($\hbar\equiv 1$ henceforth), 

\begin{align}\label{Hami}	 
    &\hat{\mathcal{H}}\left(\mathbf{r},t\right)  = -\sum_{\alpha=e,r }{\Delta}_\alpha\hat{\sigma}_{\alpha\alpha}\left(\mathbf{r},t\right)\nonumber\\
	&-\left(\frac{\Omega_{p}}{2}\hat{\sigma}_{eg}\left(\textbf{r},t\right)+\frac{\Omega_{c}}{2}\hat{\sigma}_{re}\left(\textbf{r},t\right)+\text{H.c.}\right), 
\end{align}

where $\hat{\sigma}_{\alpha\beta}$ are transition operators $(\alpha,\beta = g, e, r)$, $\Omega_p$ and $\Omega_c$ are respectively the Rabi frequencies of the probe and control fields,
${\Delta}_{e}=-\Delta_{p}+\mathbf{k}_p\cdot\mathbf{v}$ and ${\Delta}_{r}=-\Delta_{p}-\Delta_c+(\mathbf{k}_p-\mathbf{k}_c)\cdot\mathbf{v}$  are respectively the one- and two-photon detuning with $\Delta_p$ and $\Delta_c$ of the probe and coupling laser. For thermal atoms,  the detuning depends on not only the laser frequency, but also laser wave vectors and atomic velocities. As shown below, the latter two play important roles in the TRET. 
Taking into account the dissipation (e.g. atomic decay, dephasing processes, or atomic collision), the dynamics of the system is described by the master equation, $\dot{\hat{\rho}}=-i[H,\hat{\rho}]+{\mathcal L}_{\rm Decay}(\hat{\rho})+{\mathcal L}_{\rm Deph}(\hat{\rho})$, where Lindblad operators ${\mathcal L}_{\rm Decay}(\hat{\rho})$ and ${\mathcal L}_{\rm Deph}(\hat{\rho})$ describe decay and dephasing processes. The mean value $\rho_{\alpha\beta}\left(\mathbf{r},t\right) \equiv \text{Tr}\left(\hat\sigma_{\alpha\beta}\hat\rho \right)$ constitutes the optical Bloch equation.

\textit{Thermal resonance-enhanced transparency} --
To illustrate the mechanism of TRET, we evaluate integrated susceptibility of the probe light $\chi={\rm Re}(\chi)+i{\rm Im}(\chi)=2{\cal N} |\mu_{eg}|^2\int dvf(v)\rho_{ge}(v)/[\varepsilon_0\Omega_p]$, which depends on the dipole matrix element $\mu_{eg}$ between the ground state and excited state and the atomic coherence $\rho_{ge}$ (see supplemental materials (\textbf{SM}) for details), and Maxwell-Boltzmann distribution, $f(v)=1/(\sqrt{\pi}v_T){\rm exp}[-(v/v_T)^2]$. Here $v$ is the velocity projected to the laser propagation axis, and $v_T=\sqrt{2k_BT/M}$ is the
most probable atomic speed at temperature $T$ ($M$ to be mass of Cs atoms). 
Under the weak-excitation condition, where the susceptibility can be obtained analytically. Its imaginary part $\text{Im}(\chi)$ determines the absorption and is given by, 
\begin{eqnarray}\label{lineara}
		{\rm Im}(\chi)
		&\approx &\kappa\int dv\frac{\Delta_r^2f(v)}{\Gamma^2\Delta_r^2+[\Delta_e\Delta_r-|\Omega_c|^2]^2},
\end{eqnarray}
with coupling coefficient $\kappa$. 
We can obtain transmission of the probe field $T_3={\rm exp}[-k_p{\rm Im}(\chi) L]$ with medium length $L$. 
High transmission is achieved when ${\rm Im}(\chi)$ vanishes occurring when
${\Delta}_{r}=0$, i.e., $\Delta_{p}+\Delta_c=({k}_p-{k}_c)v_c$. This allows to find the center velocity $v_c$ of the atoms with given $\Delta_c$ and $\Delta_p$. For velocity groups away from $v_c$, strong absorption is expected as  ${\rm Im}(\chi)$ is non-negligible. For narrow linewidth lasers, this gives a sharp transparency window whose width is approximately $|\Omega_c|^2/\Gamma$. Though the linear approximation predicts the appearance of TRET, the transmission can only be obtained accurately by numerically solving the master equation (see \textbf{SM} for details). 

\begin{figure}[htbp]
    \centering
    \includegraphics[width=0.5\textwidth]{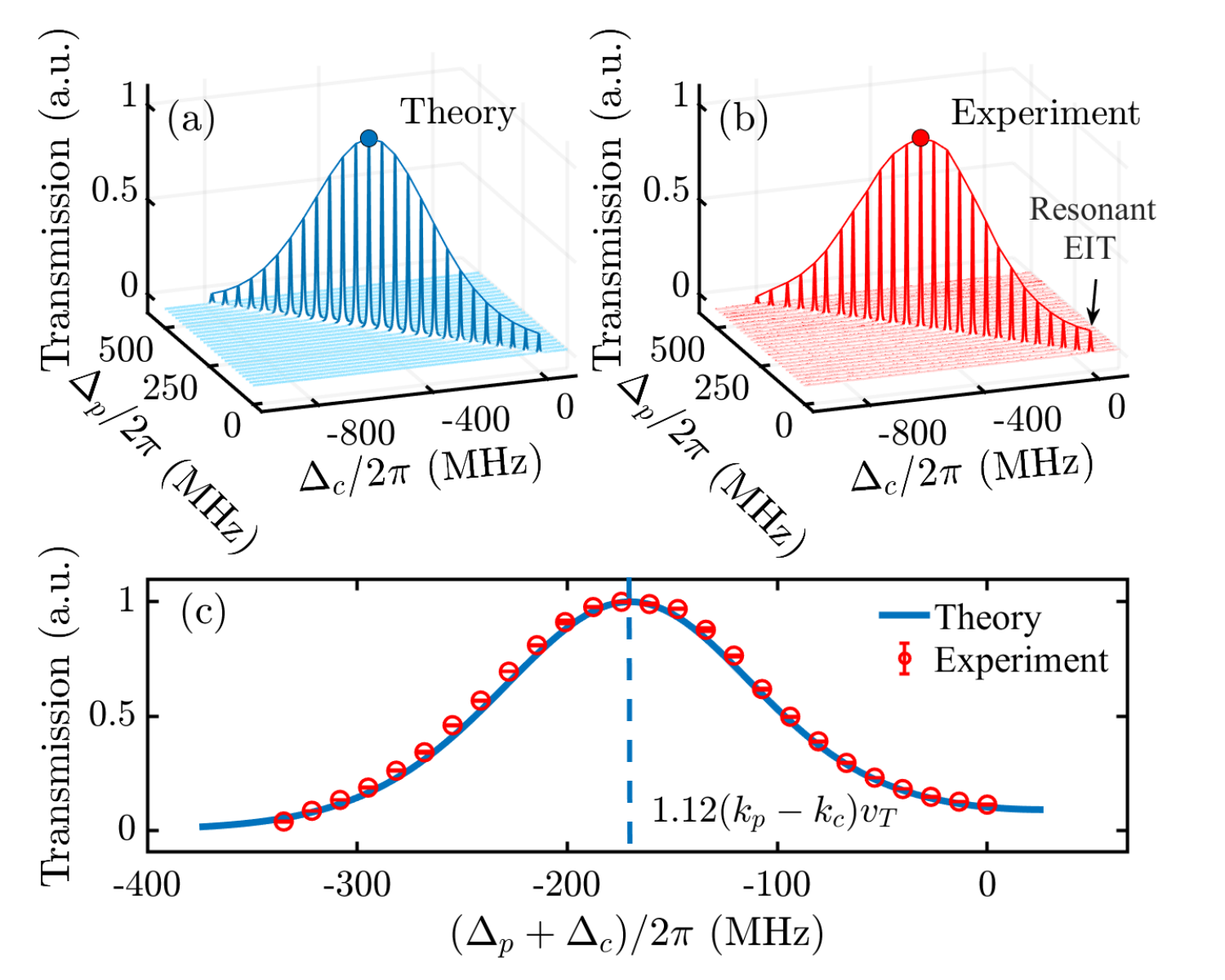}
    \caption{Transmission of the probe field. (a) Theory simulation and (b) experimental data of the transmission by varying detuning $\Delta_c$ and $\Delta_p$. The solid envelope indicates the position of thermal resonance. The transmission peak under EIT is marked. (c) The dependence of TRET peaks on the laser detuning $\Delta_p + \Delta_c$ for experimental data (red hollow circles) and simulation (blue curve). The dashed line marks the maximum transmission enhanced by the thermal resonance. The data are normalized to the maximum EIT peak. }  
    \label{figure2}
\end{figure}

Transmission for two different $v_c$ is shown in Fig.~\ref{figure1}(d) and (e).
For resonant excitation (with $v_c=0$), we obtain a transmission peak. As $\Delta_p = \Delta_c=0$, the transmission is primarily due to the EIT effect [Fig.~\ref{figure1}(d)]. We then turn to a case $v_c>0$. Here we chose $v_c$ to be the most probable speed  $v_T \simeq 193$~m/s. A transmission peak occurs at  detunings $\Delta_p=2\pi\times225$~MHz and  $\Delta_c = -\Delta_p\lambda_p/\lambda_c \simeq - 2\pi \times 377$~MHz. Surprisingly, the transmission is much higher than that of the EIT regime [Fig.~\ref{figure1}(e)]. First, to achieve the latter transmission peak, the two laser detunings are bounded, $\Delta_c=-\Delta_p\lambda_p/\lambda_c$, which results from the compensation of the single photon Doppler effect, i.e. $\Delta_p=k_p v_c$ and $\Delta_c = -k_c v_c$. Using additionally $\Delta_r=0$, we obtain $\Delta_p=2\pi v_c/\lambda_p$ and $\Delta_c = -2\pi v_c/\lambda_c$. Here only one of the three quantities $v_c$, $\Delta_p$ and $\Delta_p$ is independent. One can thus selectively excite a velocity group $v_c$ by choosing the combination of  $\Delta_p$ and $\Delta_c$ correspondingly.  Second, there is a maximal transmission when varying the laser detuning, which can be found by evaluating $\frac{\partial(T_3-T_2)}{\partial\Delta_p}=0$. Here $T_2$ is the transmission of the corresponding two-level atom. We indeed find a maximum (Fig.~\ref{figure2}(a)) at a given combination of $\Delta_p=\Delta_p^{(\text{m})}$ and $\Delta_c=\Delta_c^{(\text{m})}$. We will discuss the position of the maximum in the experimental realization section.

Moreover, the probe laser is scattered if the atomic velocity is away from $v_c$. This builds up an atomic speed filter~\cite{aumiler2005}.   Note that, refer to Eq.~(\ref{lineara}), the residual Doppler broadening due to the small wavenumber mismatch $\Delta k=k_p-k_c$ between the two laser fields is an essential ingredient to set up such TRET.
In contrast, for a three-level $\Lambda$-scheme system, the probe and control lasers typically have similar wave numbers (frequencies), where it is difficult to realize TRET.

\textit{Experiment realization of TRET} -- We perform the two-photon excitation of $^{133}$Cs  Rydberg atom experiment in a room-temperature vapor cell with a size of $\phi$ 2.5 cm $\times$ 7.5 cm. A Rydberg coupling laser ($\lambda_c=509$~nm, $\Omega_c = 2\pi \times$ 1.57~MHz ) 
and a probe laser ($\lambda_p=852$~nm, $\Omega_p = 2\pi \times$ 4.96~MHz)
counter-propagate through the cell, driving the ground state $|6S_{1/2}, F = 4\rangle$ to the Rydberg state $|52D_{5/2}\rangle$ via an intermediate state $|6P_{3/2}, F^\prime = 5\rangle$.
The probe and coupling lasers keep co-linear polarization along the $y$-axis, and their $1/e^2$ beam waists $\omega_p$ and $\omega_c$ are 800~$\mu$m and 900~$\mu$m, respectively. The probe detuning $\Delta_p$ is modified with a tunable offset locking technique using a high finesse ultralow expansion (ULE) cavity. We modulate the coupling detuning  $\Delta_c$
to constitute the atomic spectrum. A differential photodiode (DPD) is used to detect the transmission of the probe field~\cite{Carr2013}.

We excite a specific velocity group by adjusting $\Delta_p$ and $\Delta_c$ with $|\Delta_c / \Delta_p|=$ $\lambda_p / \lambda_c = 1.67$, which is the Doppler mismatch factor of the probe and coupling laser. 
Figure~\ref{figure2}(b) displays measured transmission spectra 
 as a function of coupling detuning $\Delta_c$ for the probe detuning $\Delta_p$ over a range of $0 - 500$~MHz, in steps of 20~MHz. 
When the probe laser is blue detuned to compensate the thermal-motion-induced energy shifts, we observe a series of transmission peaks. The relatively large step size ensures that individual peaks are separable. However, we can also control the peak positions by tuning the step size. 
 
Under the EIT condition $\Delta_p=\Delta_c=0$, we obtain a transmission peak (Fig.~\ref{figure2}(b)). Increasing $\Delta_p$, the height becomes higher than that the EIT case. To obtain the profile of the transmission height,  we extract the peak of the transmission spectra as a function of $\Delta_p + \Delta_c$, shown in Fig.~\ref{figure2}(c), where the red hollow circles represent the experimental data and the blue line is obtained from the simulation. It can be seen that the numerical simulation on the shape and maximal transmission of the TRET agrees nicely with the experimental data. It reaches a maximum at $\Delta_p =\Delta_p^{(\text{m})}\approx 255$~MHz. The emergence of the maximal transmission is also predicted in the theory analysis in the previous section. Importantly, the height at $\Delta_p^{(\text{m})}$ is significantly enhanced by a factor of 8.8 compared to the EIT case. Using the TRET condition, the corresponding speed can be obtained, 
 $v_c^{\rm (m)}\approx 1.12v_T$, which is much higher than the speed $v_c \approx 0$ in the EIT regime. This shows that the higher transmission is obtained for atoms with velocity $v_c^{\rm (m)}\gg 0$. Additional numerical simulations show that the relation between  $v_c^{\rm (m)}$ and $v_T$  varies with the temperature and also the laser parameters (see demonstration in \textbf{SM}).

\begin{figure}[htbp]
    \centering
    \includegraphics[width=0.5\textwidth]{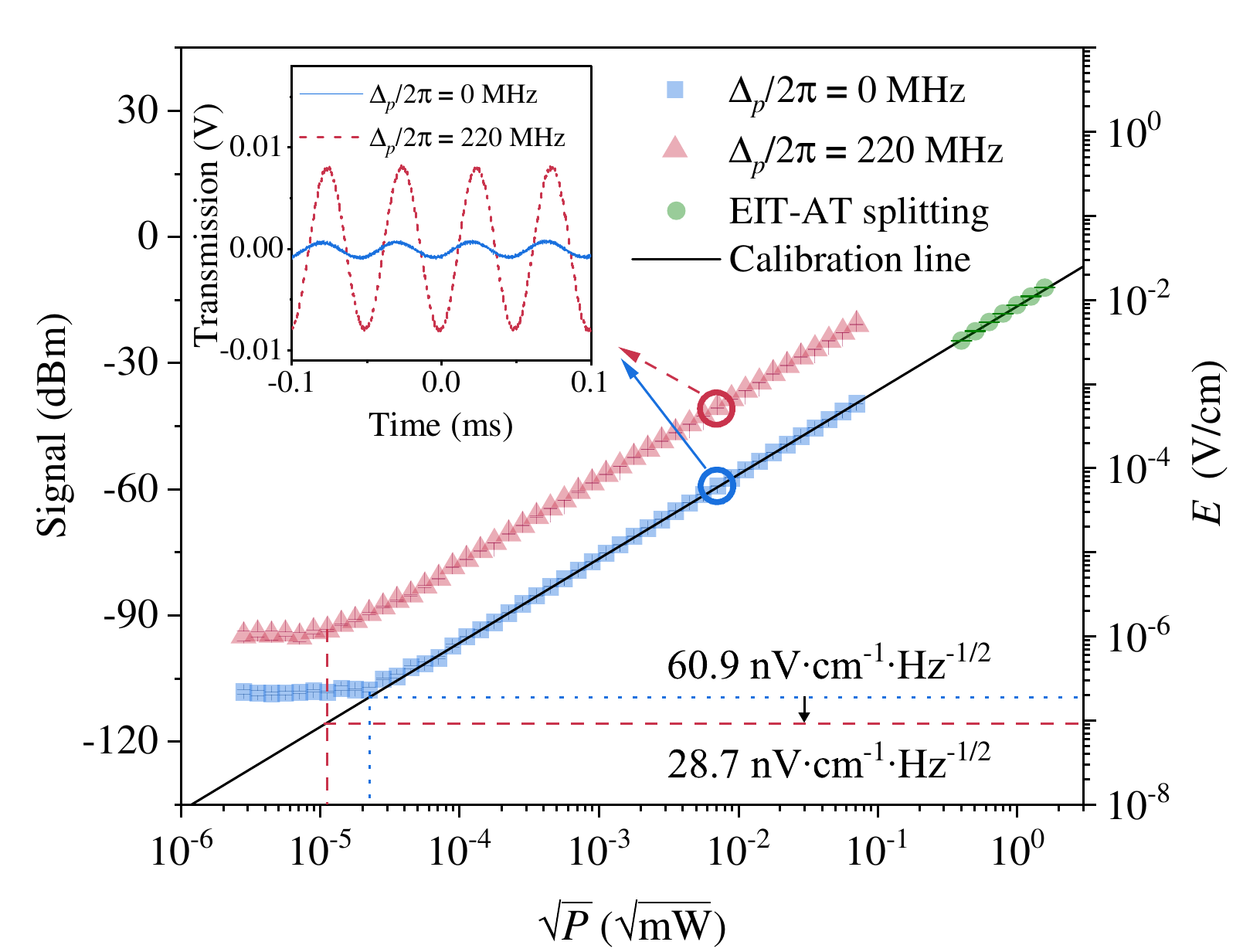}
    \caption{Measured sensitivity for $\Delta_p /2\pi$ = 0~MHz with $v_c \approx$ 0~m/s (blue squares) and 220~MHz with $v_c \approx$ 187~m/s (red triangles). The blue and red dashed lines indicate the detectable minimum electric fields. Inset shows the oscillation signal for $\Delta_p /2\pi$ = 0 (blue solid line) and 220~MHz (red dashed line) at a
    signal field $E_{Sig}=$ 60~$\mu$V/cm. The green circles show the EIT-AT
    splitting in a strong field region and the black solid line shows the calibrated electric field.}
    \label{figure3}
\end{figure}

{\it Enhancing MW field measurement sensitivity} -- %\bred{Rydberg EIT, where strong Rydberg atom interactions are mapped to light fields, finds novel applications~\cite{shaoRydbergSuperatomsArtificial2024a} in realizing single photon sources~\cite{dudin2012Strongly,peyronel2012Quantumb,ornelas-huerta2020a}, optical transistor~\cite{Hofferberth2014,tiarks2014SinglePhoton,gorniaczykEnhancementRydbergmediatedSinglephoton2016}, quantum gates~\cite{PhysRevLett.102.170502,tiarks2019Photon,Durr2022},  collectively encoded qubits~\cite{spong2021collectively}, and quantum thermometry~\cite{schlossberger2025}. 
Rydberg EIT
% In the interaction free regime,
%  Rydberg atoms 
permit sensing weak MW fields ~\cite{sedlacek2012,Jing2020,simons2021,schlossberger2024,elgee2024,robinson2021determining,holloway2014subwave,meyer2021,yuanQuantumSensingMicrowave2023,schlossberger2024a} and communications~\cite{cox2018Quantum, anderson2021, simons2019b, holloway2019}. Due to the enhanced transmission, we will show that the TRET can improve the sensitivity of the Rydberg MW sensor~\cite{prajapati2021,cai2023a,wu2025,wu2025a}. 

In our experiment, the superheterodyne technique~\cite{Jing2020} is used to measure MW field. Both local oscillator (LO) field $E_{LO}$ and signal field $E_{Sig}$ are incident to the Rydberg receiver simultaneously. The strong $E_{LO}$ resonantly couples the Rydberg transition between states $|52D_{5/2}\rangle$ and $|53P_{3/2}\rangle$ with a frequency  5.04~GHz, while the $E_{Sig}$ has a detuning of $\delta_{IF}$ = 20~kHz relative to the resonant transition. The frequency difference leads to 20~kHz oscillations of the EIT transmission.
We obtain the sensitivity of the Rydberg receiver by measuring the power of oscillation signals. In Fig.~\ref{figure3}, we demonstrate the sensitivity measurements of the Rydberg receiver with a measurement time of 0.1~s by choosing $v_c \approx$~0 and $v_c \approx$ 187~m/s, corresponding laser detuing of $\Delta_p /2\pi$ = 0 (blue squares) and 220~MHz (red triangles).
The results show that the sensitivity is improved to be 28.7~nVcm$^{-1}$Hz$^{-1/2}$ due to the TRET, which enhances by a factor of 2.1, compared with the sensitivity of 60.9~nVcm$^{-1}$Hz$^{-1/2}$ under the EIT condition. We choose $\Delta_p /2\pi =$ 220~MHz rather than 255~MHz due to its narrow EIT linewidth (see \textbf{SM}).
The inset shows the improvement of the oscillation signal for $\Delta_p /2\pi$ = 0 and 220~MHz at the same strength of signal field $E_{\rm Sig}=$ 60~$\mu$V/cm.
The EIT-AT splitting in a strong field region (green circles)
and the far-field formula $E_{\rm FF} = {F\sqrt{30P\cdot g}}/{d}$ (black solid line) are used to determine the strength of the signal fields (see \textbf{SM} for technical details).

\begin{figure*}[htbp]
    \centering
    \includegraphics[width=0.9\textwidth]{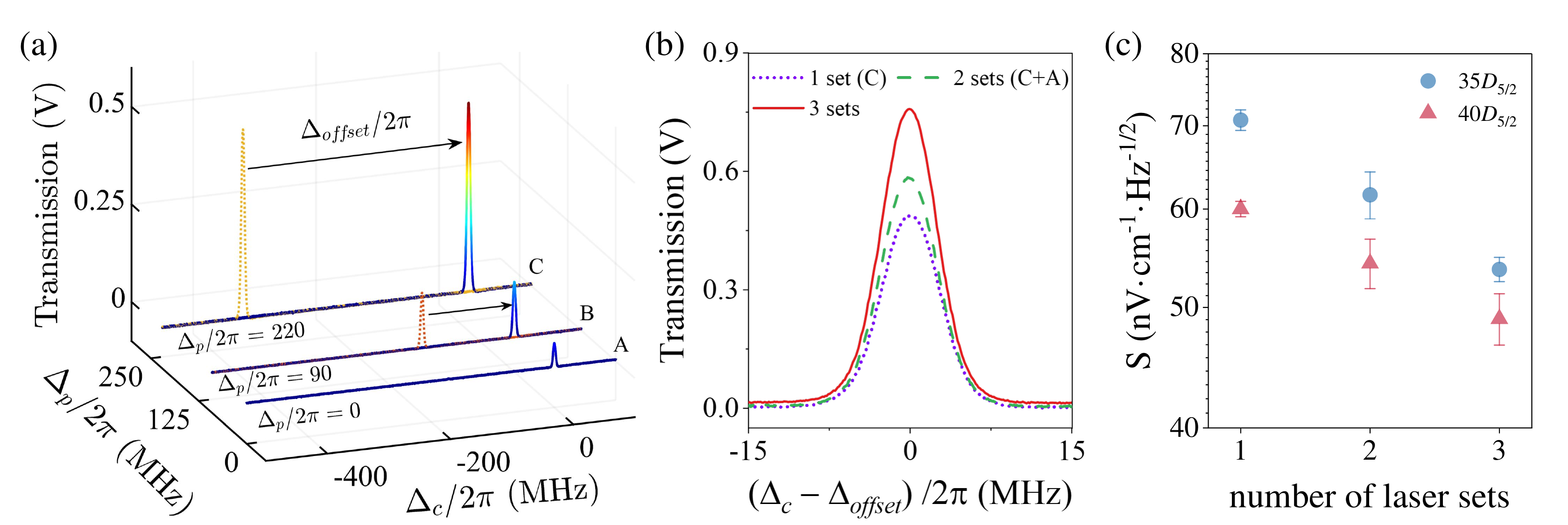}
    \caption{MW measurement with multiple sets of detuned lasers. (a) Measurement of the transmission signal of 35$D_{5/2}$ as a function of $\Delta_c$ at indicated detuning of $\Delta_p$. (b) Transmission spectra of the $35D_{5/2}$ state as a function of $\Delta_c - \Delta_{\rm offset}$ with different number of laser sets. 
    (c) Sensitivities for $35D_{5/2}$ and $40D_{5/2}$ with different sets of excitation lasers. The sensitivity shows a nearly linear enhancement with the number of laser sets.}
    \label{figure4}
\end{figure*}

As different transmission peaks are separable,  multiple sets of the detuned lasers are used to excite different velocity groups of atoms simultaneously, 
which can further improve the sensitivity.
To avoid interferences between different atom groups by the Rydberg atom interaction, a low-lying Rydberg state (principal quantum number $n=35$) is considered. We demonstrate this by employing maximally three sets of EIT excitation lasers, where the detuning of three probe lasers are 0~MHz (A), +90~MHz (B, $v_c \approx$ 77~m/s) and +220~MHz (C). The Rabi frequencies of each probe and each coupling are $\Omega_p = 2\pi \times$ 4.53~MHz and $\Omega_c = 2\pi \times$ 0.87~MHz. Fig.~\ref{figure4}(a) shows the transmission signal of 35$D_{5/2}$ as a function of $\Delta_c$ for each individual frequency set.
We see the spectra have an offset frequency $\Delta_{\rm offset}/2\pi =$ -150.3~MHz and -367.4~MHz for the probe detuning +90~MHz and +220~MHz compared with the resonance condition.  
To enhance the transmission with the three velocity groups simultaneously, we compensate the $\Delta_{\rm offset}$ by using AMOs, making three sets of transmission spectra appear in the same position (see \textbf{SM} for the details). In Fig.~\ref{figure4}(b), the transmission signal of 35$D_{5/2}$ for 1 set (C), 2 sets (C+A) and 3 sets (C+A+B) of excitation lasers can be seen. It is apparent that multiple excitation laser sets additionally facilitate the transmission signal.

Building upon this, we perform the sensitivity measurements for microwave coupling the transition of 35$D_{5/2}\to 36P_{3/2}$ with 11.62~GHz.   
The sensitivity is demonstrated as the blue circles in Fig~\ref{figure4}(c) for a single set with C (70.7~nVcm$^{-1}$Hz$^{-1/2}$), two sets with C+A (61.6~nVcm$^{-1}$Hz$^{-1/2}$) and three sets (53.6~nVcm$^{-1}$Hz$^{-1/2}$), showing the multiple sets of excitation lasers further improve the sensitivity. The enhancement of the sensitivity is linearly proportional to the number of the laser sets. Therefore, the sensitivity can be additionally improved by increasing the number of laser sets, as long as different sets will not cause interference. In addition, we perform the same experiment at 40$D_{5/2}$ (red triangles), which achieves similar improvement and linear tendency. We remark that the sensitivity at 35$D_{5/2}$ (40$D_{5/2}$) is lower than that of at 52$D_{5/2}$, as the lower microwave transition dipole moment for lower Rydberg states.

\textit{Conclusion}-- We have demonstrated the enhanced optical transmission in thermal atomic gases using TRET. The TRET is enabled when the laser fields are tuned to compensate the atomic thermal-motion-induced energy shifts. The maximal transmission is almost one order of magnitude larger than that of the EIT case. We have further demonstrated the improvement of the sensitivity of MW measurements by utilizing the TRET.
Using multiple laser sets to excite different velocity groups simultaneously,
it is found that the sensitivity of the Rydberg receiver can be enhanced even more. By tuning the probe and the coupling laser frequency to match the thermal energy shift of different atomic velocity groups, we can even obtain the velocity distribution of the atomic gas~\cite{yan2000,chen2020}, which provides a practical way to build an atomic speed filter. Our enhanced sensing scheme can be extended to sense and detect the terahertz field with lower Rydberg states~\cite{wade2017a,downes2020b}.

\textit{Acknowledgment}-- The work is supported by the National Natural Science Foundation of China (No. U2341211, 62175136, 12274131, 12241408, 12120101004); Innovation Program for Quantum Science and Technology (No. 2023ZD0300902, 2024ZD0300101);  Fundamental Research Program of Shanxi Province (No. 202303021224007); and the 1331 project of Shanxi Province. W.L. acknowledges financial support from the EPSRC (Grant No. EP/W015641/1) and the Going Global Partnerships Programme of the British Council (Contract No. IND/CONT/G/22-23/26).
 
% \section*{Data Availability Statement}
% The data that support the findings of this study are available from the corresponding author upon reasonable request.
	
% \section{References}
\bibliography{main}% Produces the bibliography via BibTeX.

\begin{thebibliography}{50}%
\makeatletter
\providecommand \@ifxundefined [1]{%
 \@ifx{#1\undefined}
}%
\providecommand \@ifnum [1]{%
 \ifnum #1\expandafter \@firstoftwo
 \else \expandafter \@secondoftwo
 \fi
}%
\providecommand \@ifx [1]{%
 \ifx #1\expandafter \@firstoftwo
 \else \expandafter \@secondoftwo
 \fi
}%
\providecommand \natexlab [1]{#1}%
\providecommand \enquote  [1]{``#1''}%
\providecommand \bibnamefont  [1]{#1}%
\providecommand \bibfnamefont [1]{#1}%
\providecommand \citenamefont [1]{#1}%
\providecommand \href@noop [0]{\@secondoftwo}%
\providecommand \href [0]{\begingroup \@sanitize@url \@href}%
\providecommand \@href[1]{\@@startlink{#1}\@@href}%
\providecommand \@@href[1]{\endgroup#1\@@endlink}%
\providecommand \@sanitize@url [0]{\catcode `\\12\catcode `\$12\catcode
  `\&12\catcode `\#12\catcode `\^12\catcode `\_12\catcode `\%12\relax}%
\providecommand \@@startlink[1]{}%
\providecommand \@@endlink[0]{}%
\providecommand \url  [0]{\begingroup\@sanitize@url \@url }%
\providecommand \@url [1]{\endgroup\@href {#1}{\urlprefix }}%
\providecommand \urlprefix  [0]{URL }%
\providecommand \Eprint [0]{\href }%
\providecommand \doibase [0]{https://doi.org/}%
\providecommand \selectlanguage [0]{\@gobble}%
\providecommand \bibinfo  [0]{\@secondoftwo}%
\providecommand \bibfield  [0]{\@secondoftwo}%
\providecommand \translation [1]{[#1]}%
\providecommand \BibitemOpen [0]{}%
\providecommand \bibitemStop [0]{}%
\providecommand \bibitemNoStop [0]{.\EOS\space}%
\providecommand \EOS [0]{\spacefactor3000\relax}%
\providecommand \BibitemShut  [1]{\csname bibitem#1\endcsname}%
\let\auto@bib@innerbib\@empty
%</preamble>
\bibitem [{\citenamefont {Fleischhauer}\ \emph
  {et~al.}(2005{\natexlab{a}})\citenamefont {Fleischhauer}, \citenamefont
  {Imamoglu},\ and\ \citenamefont {Marangos}}]{RevModPhys.77.633}%
  \BibitemOpen
  \bibfield  {author} {\bibinfo {author} {\bibfnamefont {M.}~\bibnamefont
  {Fleischhauer}}, \bibinfo {author} {\bibfnamefont {A.}~\bibnamefont
  {Imamoglu}},\ and\ \bibinfo {author} {\bibfnamefont {J.~P.}\ \bibnamefont
  {Marangos}},\ }\bibfield  {title} {\bibinfo {title} {Electromagnetically
  induced transparency: Optics in coherent media},\ }\href
  {https://doi.org/10.1103/RevModPhys.77.633} {\bibfield  {journal} {\bibinfo
  {journal} {Reviews of Modern Physics}\ }\textbf {\bibinfo {volume} {77}},\
  \bibinfo {pages} {633} (\bibinfo {year} {2005}{\natexlab{a}})}\BibitemShut
  {NoStop}%
\bibitem [{\citenamefont {Saffman}\ \emph {et~al.}(2010)\citenamefont
  {Saffman}, \citenamefont {Walker},\ and\ \citenamefont
  {M\o{}lmer}}]{saffman2010}%
  \BibitemOpen
  \bibfield  {author} {\bibinfo {author} {\bibfnamefont {M.}~\bibnamefont
  {Saffman}}, \bibinfo {author} {\bibfnamefont {T.~G.}\ \bibnamefont
  {Walker}},\ and\ \bibinfo {author} {\bibfnamefont {K.}~\bibnamefont
  {M\o{}lmer}},\ }\bibfield  {title} {\bibinfo {title} {Quantum information
  with rydberg atoms},\ }\href {https://doi.org/10.1103/RevModPhys.82.2313}
  {\bibfield  {journal} {\bibinfo  {journal} {Reviews of Modern Physics}\
  }\textbf {\bibinfo {volume} {82}},\ \bibinfo {pages} {2313} (\bibinfo {year}
  {2010})}\BibitemShut {NoStop}%
\bibitem [{\citenamefont {Noh}\ and\ \citenamefont
  {Angelakis}(2016)}]{nohQuantumSimulationsManybody2016}%
  \BibitemOpen
  \bibfield  {author} {\bibinfo {author} {\bibfnamefont {C.}~\bibnamefont
  {Noh}}\ and\ \bibinfo {author} {\bibfnamefont {D.~G.}\ \bibnamefont
  {Angelakis}},\ }\bibfield  {title} {\bibinfo {title} {Quantum simulations and
  many-body physics with light},\ }\href
  {https://doi.org/10.1088/0034-4885/80/1/016401} {\bibfield  {journal}
  {\bibinfo  {journal} {Reports On Progress In Physics}\ }\textbf {\bibinfo
  {volume} {80}},\ \bibinfo {pages} {016401} (\bibinfo {year}
  {2016})}\BibitemShut {NoStop}%
\bibitem [{\citenamefont {Adams}\ \emph {et~al.}(2019)\citenamefont {Adams},
  \citenamefont {Pritchard},\ and\ \citenamefont {Shaffer}}]{adams2019}%
  \BibitemOpen
  \bibfield  {author} {\bibinfo {author} {\bibfnamefont {C.~S.}\ \bibnamefont
  {Adams}}, \bibinfo {author} {\bibfnamefont {J.~D.}\ \bibnamefont
  {Pritchard}},\ and\ \bibinfo {author} {\bibfnamefont {J.~P.}\ \bibnamefont
  {Shaffer}},\ }\bibfield  {title} {\bibinfo {title} {Rydberg atom quantum
  technologies},\ }\href {https://doi.org/10.1088/1361-6455/ab52ef} {\bibfield
  {journal} {\bibinfo  {journal} {Journal of Physics B: Atomic, Molecular and
  Optical Physics}\ }\textbf {\bibinfo {volume} {53}},\ \bibinfo {pages}
  {012002} (\bibinfo {year} {2019})}\BibitemShut {NoStop}%
\bibitem [{\citenamefont {Aoki}\ \emph {et~al.}(2006)\citenamefont {Aoki},
  \citenamefont {Dayan}, \citenamefont {Wilcut}, \citenamefont {Bowen},
  \citenamefont {Parkins}, \citenamefont {Kippenberg}, \citenamefont {Vahala},\
  and\ \citenamefont {Kimble}}]{aoki2006}%
  \BibitemOpen
  \bibfield  {author} {\bibinfo {author} {\bibfnamefont {T.}~\bibnamefont
  {Aoki}}, \bibinfo {author} {\bibfnamefont {B.}~\bibnamefont {Dayan}},
  \bibinfo {author} {\bibfnamefont {E.}~\bibnamefont {Wilcut}}, \bibinfo
  {author} {\bibfnamefont {W.~P.}\ \bibnamefont {Bowen}}, \bibinfo {author}
  {\bibfnamefont {A.~S.}\ \bibnamefont {Parkins}}, \bibinfo {author}
  {\bibfnamefont {T.~J.}\ \bibnamefont {Kippenberg}}, \bibinfo {author}
  {\bibfnamefont {K.~J.}\ \bibnamefont {Vahala}},\ and\ \bibinfo {author}
  {\bibfnamefont {H.~J.}\ \bibnamefont {Kimble}},\ }\bibfield  {title}
  {\bibinfo {title} {Observation of strong coupling between one atom and a
  monolithic microresonator},\ }\href {https://doi.org/10.1038/nature05147}
  {\bibfield  {journal} {\bibinfo  {journal} {Nature}\ }\textbf {\bibinfo
  {volume} {443}},\ \bibinfo {pages} {671} (\bibinfo {year}
  {2006})}\BibitemShut {NoStop}%
\bibitem [{\citenamefont {Lepert}\ \emph {et~al.}(2011)\citenamefont {Lepert},
  \citenamefont {Trupke}, \citenamefont {Hartmann}, \citenamefont {Plenio},\
  and\ \citenamefont {Hinds}}]{lepert2011}%
  \BibitemOpen
  \bibfield  {author} {\bibinfo {author} {\bibfnamefont {G.}~\bibnamefont
  {Lepert}}, \bibinfo {author} {\bibfnamefont {M.}~\bibnamefont {Trupke}},
  \bibinfo {author} {\bibfnamefont {M.~J.}\ \bibnamefont {Hartmann}}, \bibinfo
  {author} {\bibfnamefont {M.~B.}\ \bibnamefont {Plenio}},\ and\ \bibinfo
  {author} {\bibfnamefont {E.~A.}\ \bibnamefont {Hinds}},\ }\bibfield  {title}
  {\bibinfo {title} {Arrays of waveguide-coupled optical cavities that interact
  strongly with atoms},\ }\href
  {https://doi.org/10.1088/1367-2630/13/11/113002} {\bibfield  {journal}
  {\bibinfo  {journal} {New Journal of Physics}\ }\textbf {\bibinfo {volume}
  {13}},\ \bibinfo {pages} {113002} (\bibinfo {year} {2011})}\BibitemShut
  {NoStop}%
\bibitem [{\citenamefont {Bettles}\ \emph {et~al.}(2016)\citenamefont
  {Bettles}, \citenamefont {Gardiner},\ and\ \citenamefont
  {Adams}}]{bettles2016}%
  \BibitemOpen
  \bibfield  {author} {\bibinfo {author} {\bibfnamefont {R.~J.}\ \bibnamefont
  {Bettles}}, \bibinfo {author} {\bibfnamefont {S.~A.}\ \bibnamefont
  {Gardiner}},\ and\ \bibinfo {author} {\bibfnamefont {C.~S.}\ \bibnamefont
  {Adams}},\ }\bibfield  {title} {\bibinfo {title} {Enhanced {{Optical Cross
  Section}} via {{Collective Coupling}} of {{Atomic Dipoles}} in a {{2D
  Array}}},\ }\href {https://doi.org/10.1103/PhysRevLett.116.103602} {\bibfield
   {journal} {\bibinfo  {journal} {Physical Review Letters}\ }\textbf {\bibinfo
  {volume} {116}},\ \bibinfo {pages} {103602} (\bibinfo {year}
  {2016})}\BibitemShut {NoStop}%
\bibitem [{\citenamefont {Ferioli}\ \emph {et~al.}(2021)\citenamefont
  {Ferioli}, \citenamefont {Glicenstein}, \citenamefont {Henriet},
  \citenamefont {{Ferrier-Barbut}},\ and\ \citenamefont
  {Browaeys}}]{ferioli2021}%
  \BibitemOpen
  \bibfield  {author} {\bibinfo {author} {\bibfnamefont {G.}~\bibnamefont
  {Ferioli}}, \bibinfo {author} {\bibfnamefont {A.}~\bibnamefont
  {Glicenstein}}, \bibinfo {author} {\bibfnamefont {L.}~\bibnamefont
  {Henriet}}, \bibinfo {author} {\bibfnamefont {I.}~\bibnamefont
  {{Ferrier-Barbut}}},\ and\ \bibinfo {author} {\bibfnamefont {A.}~\bibnamefont
  {Browaeys}},\ }\bibfield  {title} {\bibinfo {title} {Storage and {{Release}}
  of {{Subradiant Excitations}} in a {{Dense Atomic Cloud}}},\ }\href
  {https://doi.org/10.1103/PhysRevX.11.021031} {\bibfield  {journal} {\bibinfo
  {journal} {Physical Review X}\ }\textbf {\bibinfo {volume} {11}},\ \bibinfo
  {pages} {021031} (\bibinfo {year} {2021})}\BibitemShut {NoStop}%
\bibitem [{\citenamefont {Boller}\ \emph {et~al.}(1991)\citenamefont {Boller},
  \citenamefont {Imamo{\u g}lu},\ and\ \citenamefont {Harris}}]{boller1991}%
  \BibitemOpen
  \bibfield  {author} {\bibinfo {author} {\bibfnamefont {K.-J.}\ \bibnamefont
  {Boller}}, \bibinfo {author} {\bibfnamefont {A.}~\bibnamefont {Imamo{\u
  g}lu}},\ and\ \bibinfo {author} {\bibfnamefont {S.~E.}\ \bibnamefont
  {Harris}},\ }\bibfield  {title} {\bibinfo {title} {Observation of
  electromagnetically induced transparency},\ }\href
  {https://doi.org/10.1103/PhysRevLett.66.2593} {\bibfield  {journal} {\bibinfo
   {journal} {Physical Review Letters}\ }\textbf {\bibinfo {volume} {66}},\
  \bibinfo {pages} {2593} (\bibinfo {year} {1991})}\BibitemShut {NoStop}%
\bibitem [{\citenamefont {Fleischhauer}\ \emph
  {et~al.}(2005{\natexlab{b}})\citenamefont {Fleischhauer}, \citenamefont
  {Imamoglu},\ and\ \citenamefont {Marangos}}]{fleischhauer2005a}%
  \BibitemOpen
  \bibfield  {author} {\bibinfo {author} {\bibfnamefont {M.}~\bibnamefont
  {Fleischhauer}}, \bibinfo {author} {\bibfnamefont {A.}~\bibnamefont
  {Imamoglu}},\ and\ \bibinfo {author} {\bibfnamefont {J.~P.}\ \bibnamefont
  {Marangos}},\ }\bibfield  {title} {\bibinfo {title} {Electromagnetically
  induced transparency: {{Optics}} in coherent media},\ }\href
  {https://doi.org/10.1103/RevModPhys.77.633} {\bibfield  {journal} {\bibinfo
  {journal} {Reviews of Modern Physics}\ }\textbf {\bibinfo {volume} {77}},\
  \bibinfo {pages} {633} (\bibinfo {year} {2005}{\natexlab{b}})}\BibitemShut
  {NoStop}%
\bibitem [{\citenamefont {Chu}\ \emph {et~al.}(2025)\citenamefont {Chu},
  \citenamefont {Lu}, \citenamefont {Chiang}, \citenamefont {Lin},
  \citenamefont {Chen}, \citenamefont {Yu}, \citenamefont {Liao},\ and\
  \citenamefont {Chen}}]{chu2025}%
  \BibitemOpen
  \bibfield  {author} {\bibinfo {author} {\bibfnamefont {K.-I.}\ \bibnamefont
  {Chu}}, \bibinfo {author} {\bibfnamefont {X.-C.}\ \bibnamefont {Lu}},
  \bibinfo {author} {\bibfnamefont {K.-H.}\ \bibnamefont {Chiang}}, \bibinfo
  {author} {\bibfnamefont {Y.-H.}\ \bibnamefont {Lin}}, \bibinfo {author}
  {\bibfnamefont {C.-D.}\ \bibnamefont {Chen}}, \bibinfo {author}
  {\bibfnamefont {I.~A.}\ \bibnamefont {Yu}}, \bibinfo {author} {\bibfnamefont
  {W.-T.}\ \bibnamefont {Liao}},\ and\ \bibinfo {author} {\bibfnamefont
  {Y.-F.}\ \bibnamefont {Chen}},\ }\bibfield  {title} {\bibinfo {title} {Slow
  and stored light via electromagnetically induced transparency using a
  {{$\Lambda$}} -type superconducting artificial atom},\ }\href
  {https://doi.org/10.1103/PhysRevResearch.7.L012015} {\bibfield  {journal}
  {\bibinfo  {journal} {Physical Review Research}\ }\textbf {\bibinfo {volume}
  {7}},\ \bibinfo {pages} {L012015} (\bibinfo {year} {2025})}\BibitemShut
  {NoStop}%
\bibitem [{\citenamefont {Liu}\ \emph {et~al.}(2021)\citenamefont {Liu},
  \citenamefont {Wei}, \citenamefont {Li}, \citenamefont {Yu}, \citenamefont
  {Liu}, \citenamefont {Yu},\ and\ \citenamefont {Wang}}]{liu2021}%
  \BibitemOpen
  \bibfield  {author} {\bibinfo {author} {\bibfnamefont {S.}~\bibnamefont
  {Liu}}, \bibinfo {author} {\bibfnamefont {Y.}~\bibnamefont {Wei}}, \bibinfo
  {author} {\bibfnamefont {X.}~\bibnamefont {Li}}, \bibinfo {author}
  {\bibfnamefont {Y.}~\bibnamefont {Yu}}, \bibinfo {author} {\bibfnamefont
  {J.}~\bibnamefont {Liu}}, \bibinfo {author} {\bibfnamefont {S.}~\bibnamefont
  {Yu}},\ and\ \bibinfo {author} {\bibfnamefont {X.}~\bibnamefont {Wang}},\
  }\bibfield  {title} {\bibinfo {title} {Dual-resonance enhanced quantum
  light-matter interactions in deterministically coupled
  quantum-dot-micropillars},\ }\href
  {https://doi.org/10.1038/s41377-021-00604-8} {\bibfield  {journal} {\bibinfo
  {journal} {Light: Science \& Applications}\ }\textbf {\bibinfo {volume}
  {10}},\ \bibinfo {pages} {158} (\bibinfo {year} {2021})}\BibitemShut
  {NoStop}%
\bibitem [{\citenamefont {Allen}\ and\ \citenamefont
  {Eberly}(1975)}]{allen2012}%
  \BibitemOpen
  \bibfield  {author} {\bibinfo {author} {\bibfnamefont {L.}~\bibnamefont
  {Allen}}\ and\ \bibinfo {author} {\bibfnamefont {J.~H.}\ \bibnamefont
  {Eberly}},\ }\href {https://www.osti.gov/biblio/7365050} {\emph {\bibinfo
  {title} {Optical resonance and two-level atoms}}}\ (\bibinfo  {publisher}
  {John Wiley and Sons, Inc., New York},\ \bibinfo {year} {1975})\BibitemShut
  {NoStop}%
\bibitem [{\citenamefont {Abi-Salloum}(2010)}]{abi2010}%
  \BibitemOpen
  \bibfield  {author} {\bibinfo {author} {\bibfnamefont {T.~Y.}\ \bibnamefont
  {Abi-Salloum}},\ }\bibfield  {title} {\bibinfo {title} {Electromagnetically
  induced transparency and autler-townes splitting: Two similar but distinct
  phenomena in two categories of three-level atomic systems},\ }\href
  {https://doi.org/10.1103/PhysRevA.81.053836} {\bibfield  {journal} {\bibinfo
  {journal} {Physical Review A}\ }\textbf {\bibinfo {volume} {81}},\ \bibinfo
  {pages} {053836} (\bibinfo {year} {2010})}\BibitemShut {NoStop}%
\bibitem [{\citenamefont {Phillips}\ \emph {et~al.}(2001)\citenamefont
  {Phillips}, \citenamefont {Fleischhauer}, \citenamefont {Mair}, \citenamefont
  {Walsworth},\ and\ \citenamefont {Lukin}}]{phillips2001}%
  \BibitemOpen
  \bibfield  {author} {\bibinfo {author} {\bibfnamefont {D.~F.}\ \bibnamefont
  {Phillips}}, \bibinfo {author} {\bibfnamefont {A.}~\bibnamefont
  {Fleischhauer}}, \bibinfo {author} {\bibfnamefont {A.}~\bibnamefont {Mair}},
  \bibinfo {author} {\bibfnamefont {R.~L.}\ \bibnamefont {Walsworth}},\ and\
  \bibinfo {author} {\bibfnamefont {M.~D.}\ \bibnamefont {Lukin}},\ }\bibfield
  {title} {\bibinfo {title} {Storage of {{Light}} in {{Atomic Vapor}}},\ }\href
  {https://doi.org/10.1103/PhysRevLett.86.783} {\bibfield  {journal} {\bibinfo
  {journal} {Physical Review Letters}\ }\textbf {\bibinfo {volume} {86}},\
  \bibinfo {pages} {783} (\bibinfo {year} {2001})}\BibitemShut {NoStop}%
\bibitem [{\citenamefont {{Gea-Banacloche}}\ \emph {et~al.}(1995)\citenamefont
  {{Gea-Banacloche}}, \citenamefont {Li}, \citenamefont {Jin},\ and\
  \citenamefont {Xiao}}]{gea-banacloche1995a}%
  \BibitemOpen
  \bibfield  {author} {\bibinfo {author} {\bibfnamefont {J.}~\bibnamefont
  {{Gea-Banacloche}}}, \bibinfo {author} {\bibfnamefont {Y.-q.}\ \bibnamefont
  {Li}}, \bibinfo {author} {\bibfnamefont {S.-z.}\ \bibnamefont {Jin}},\ and\
  \bibinfo {author} {\bibfnamefont {M.}~\bibnamefont {Xiao}},\ }\bibfield
  {title} {\bibinfo {title} {Electromagnetically induced transparency in
  ladder-type inhomogeneously broadened media: {{Theory}} and experiment},\
  }\href {https://doi.org/10.1103/PhysRevA.51.576} {\bibfield  {journal}
  {\bibinfo  {journal} {Physical Review A}\ }\textbf {\bibinfo {volume} {51}},\
  \bibinfo {pages} {576} (\bibinfo {year} {1995})}\BibitemShut {NoStop}%
\bibitem [{\citenamefont {Wu}\ \emph {et~al.}(2008)\citenamefont {Wu},
  \citenamefont {{Gea-Banacloche}},\ and\ \citenamefont {Xiao}}]{wu2008}%
  \BibitemOpen
  \bibfield  {author} {\bibinfo {author} {\bibfnamefont {H.}~\bibnamefont
  {Wu}}, \bibinfo {author} {\bibfnamefont {J.}~\bibnamefont
  {{Gea-Banacloche}}},\ and\ \bibinfo {author} {\bibfnamefont {M.}~\bibnamefont
  {Xiao}},\ }\bibfield  {title} {\bibinfo {title} {Observation of {{Intracavity
  Electromagnetically Induced Transparency}} and {{Polariton Resonances}} in a
  {{Doppler-Broadened Medium}}},\ }\href
  {https://doi.org/10.1103/PhysRevLett.100.173602} {\bibfield  {journal}
  {\bibinfo  {journal} {Physical Review Letters}\ }\textbf {\bibinfo {volume}
  {100}},\ \bibinfo {pages} {173602} (\bibinfo {year} {2008})}\BibitemShut
  {NoStop}%
\bibitem [{\citenamefont {Urvoy}\ \emph {et~al.}(2015)\citenamefont {Urvoy},
  \citenamefont {Ripka}, \citenamefont {Lesanovsky}, \citenamefont {Booth},
  \citenamefont {Shaffer}, \citenamefont {Pfau},\ and\ \citenamefont
  {L{\"o}w}}]{urvoy2015}%
  \BibitemOpen
  \bibfield  {author} {\bibinfo {author} {\bibfnamefont {A.}~\bibnamefont
  {Urvoy}}, \bibinfo {author} {\bibfnamefont {F.}~\bibnamefont {Ripka}},
  \bibinfo {author} {\bibfnamefont {I.}~\bibnamefont {Lesanovsky}}, \bibinfo
  {author} {\bibfnamefont {D.}~\bibnamefont {Booth}}, \bibinfo {author}
  {\bibfnamefont {J.~P.}\ \bibnamefont {Shaffer}}, \bibinfo {author}
  {\bibfnamefont {T.}~\bibnamefont {Pfau}},\ and\ \bibinfo {author}
  {\bibfnamefont {R.}~\bibnamefont {L{\"o}w}},\ }\bibfield  {title} {\bibinfo
  {title} {Strongly {{Correlated Growth}} of {{Rydberg Aggregates}} in a
  {{Vapor Cell}}},\ }\href {https://doi.org/10.1103/PhysRevLett.114.203002}
  {\bibfield  {journal} {\bibinfo  {journal} {Physical Review Letters}\
  }\textbf {\bibinfo {volume} {114}},\ \bibinfo {pages} {203002} (\bibinfo
  {year} {2015})}\BibitemShut {NoStop}%
\bibitem [{\citenamefont {Zhang}\ and\ \citenamefont
  {Evers}(2016)}]{zhang2016a}%
  \BibitemOpen
  \bibfield  {author} {\bibinfo {author} {\bibfnamefont {L.}~\bibnamefont
  {Zhang}}\ and\ \bibinfo {author} {\bibfnamefont {J.}~\bibnamefont {Evers}},\
  }\bibfield  {title} {\bibinfo {title} {Nonlocal nonlinear response of thermal
  {{Rydberg}} atoms and modulational instability in an absorptive nonlinear
  medium},\ }\href {https://doi.org/10.1103/PhysRevA.94.033402} {\bibfield
  {journal} {\bibinfo  {journal} {Physical Review A}\ }\textbf {\bibinfo
  {volume} {94}},\ \bibinfo {pages} {033402} (\bibinfo {year}
  {2016})}\BibitemShut {NoStop}%
\bibitem [{\citenamefont {Baluktsian}\ \emph {et~al.}(2013)\citenamefont
  {Baluktsian}, \citenamefont {Huber}, \citenamefont {L{\"o}w},\ and\
  \citenamefont {Pfau}}]{baluktsian2013}%
  \BibitemOpen
  \bibfield  {author} {\bibinfo {author} {\bibfnamefont {T.}~\bibnamefont
  {Baluktsian}}, \bibinfo {author} {\bibfnamefont {B.}~\bibnamefont {Huber}},
  \bibinfo {author} {\bibfnamefont {R.}~\bibnamefont {L{\"o}w}},\ and\ \bibinfo
  {author} {\bibfnamefont {T.}~\bibnamefont {Pfau}},\ }\bibfield  {title}
  {\bibinfo {title} {Evidence for {{Strong}} van der {{Waals Type
  Rydberg-Rydberg Interaction}} in a {{Thermal Vapor}}},\ }\href
  {https://doi.org/10.1103/PhysRevLett.110.123001} {\bibfield  {journal}
  {\bibinfo  {journal} {Physical Review Letters}\ }\textbf {\bibinfo {volume}
  {110}},\ \bibinfo {pages} {123001} (\bibinfo {year} {2013})}\BibitemShut
  {NoStop}%
\bibitem [{\citenamefont {Bai}\ \emph {et~al.}(2020)\citenamefont {Bai},
  \citenamefont {Adams}, \citenamefont {Huang},\ and\ \citenamefont
  {Li}}]{bai2020}%
  \BibitemOpen
  \bibfield  {author} {\bibinfo {author} {\bibfnamefont {Z.}~\bibnamefont
  {Bai}}, \bibinfo {author} {\bibfnamefont {C.~S.}\ \bibnamefont {Adams}},
  \bibinfo {author} {\bibfnamefont {G.}~\bibnamefont {Huang}},\ and\ \bibinfo
  {author} {\bibfnamefont {W.}~\bibnamefont {Li}},\ }\bibfield  {title}
  {\bibinfo {title} {Self-{{Induced Transparency}} in {{Warm}} and {{Strongly
  Interacting Rydberg Gases}}},\ }\href
  {https://doi.org/10.1103/PhysRevLett.125.263605} {\bibfield  {journal}
  {\bibinfo  {journal} {Physical Review Letters}\ }\textbf {\bibinfo {volume}
  {125}},\ \bibinfo {pages} {263605} (\bibinfo {year} {2020})}\BibitemShut
  {NoStop}%
\bibitem [{\citenamefont {Zhang}\ \emph {et~al.}(2018)\citenamefont {Zhang},
  \citenamefont {Hu}, \citenamefont {Lin}, \citenamefont {Niu}, \citenamefont
  {Xia}, \citenamefont {Gong},\ and\ \citenamefont {Gong}}]{zhang2018a}%
  \BibitemOpen
  \bibfield  {author} {\bibinfo {author} {\bibfnamefont {S.}~\bibnamefont
  {Zhang}}, \bibinfo {author} {\bibfnamefont {Y.}~\bibnamefont {Hu}}, \bibinfo
  {author} {\bibfnamefont {G.}~\bibnamefont {Lin}}, \bibinfo {author}
  {\bibfnamefont {Y.}~\bibnamefont {Niu}}, \bibinfo {author} {\bibfnamefont
  {K.}~\bibnamefont {Xia}}, \bibinfo {author} {\bibfnamefont {J.}~\bibnamefont
  {Gong}},\ and\ \bibinfo {author} {\bibfnamefont {S.}~\bibnamefont {Gong}},\
  }\bibfield  {title} {\bibinfo {title} {Thermal-motion-induced non-reciprocal
  quantum optical system},\ }\href {https://doi.org/10.1038/s41566-018-0269-2}
  {\bibfield  {journal} {\bibinfo  {journal} {Nature Photonics}\ }\textbf
  {\bibinfo {volume} {12}},\ \bibinfo {pages} {744} (\bibinfo {year}
  {2018})}\BibitemShut {NoStop}%
\bibitem [{\citenamefont {Liang}\ \emph {et~al.}(2020)\citenamefont {Liang},
  \citenamefont {Liu}, \citenamefont {Xu}, \citenamefont {Wen}, \citenamefont
  {Lu}, \citenamefont {Xia}, \citenamefont {Tey}, \citenamefont {Liu},\ and\
  \citenamefont {You}}]{liang2020}%
  \BibitemOpen
  \bibfield  {author} {\bibinfo {author} {\bibfnamefont {C.}~\bibnamefont
  {Liang}}, \bibinfo {author} {\bibfnamefont {B.}~\bibnamefont {Liu}}, \bibinfo
  {author} {\bibfnamefont {A.-N.}\ \bibnamefont {Xu}}, \bibinfo {author}
  {\bibfnamefont {X.}~\bibnamefont {Wen}}, \bibinfo {author} {\bibfnamefont
  {C.}~\bibnamefont {Lu}}, \bibinfo {author} {\bibfnamefont {K.}~\bibnamefont
  {Xia}}, \bibinfo {author} {\bibfnamefont {M.~K.}\ \bibnamefont {Tey}},
  \bibinfo {author} {\bibfnamefont {Y.-C.}\ \bibnamefont {Liu}},\ and\ \bibinfo
  {author} {\bibfnamefont {L.}~\bibnamefont {You}},\ }\bibfield  {title}
  {\bibinfo {title} {Collision-{{Induced Broadband Optical Nonreciprocity}}},\
  }\href {https://doi.org/10.1103/PhysRevLett.125.123901} {\bibfield  {journal}
  {\bibinfo  {journal} {Physical Review Letters}\ }\textbf {\bibinfo {volume}
  {125}},\ \bibinfo {pages} {123901} (\bibinfo {year} {2020})}\BibitemShut
  {NoStop}%
\bibitem [{\citenamefont {Dong}\ \emph {et~al.}(2021)\citenamefont {Dong},
  \citenamefont {Xia}, \citenamefont {Zhang}, \citenamefont {Yu}, \citenamefont
  {Ye}, \citenamefont {Li}, \citenamefont {Zeng}, \citenamefont {Ding},
  \citenamefont {Shi}, \citenamefont {Guo},\ and\ \citenamefont
  {Nori}}]{dong2021}%
  \BibitemOpen
  \bibfield  {author} {\bibinfo {author} {\bibfnamefont {M.-X.}\ \bibnamefont
  {Dong}}, \bibinfo {author} {\bibfnamefont {K.-Y.}\ \bibnamefont {Xia}},
  \bibinfo {author} {\bibfnamefont {W.-H.}\ \bibnamefont {Zhang}}, \bibinfo
  {author} {\bibfnamefont {Y.-C.}\ \bibnamefont {Yu}}, \bibinfo {author}
  {\bibfnamefont {Y.-H.}\ \bibnamefont {Ye}}, \bibinfo {author} {\bibfnamefont
  {E.-Z.}\ \bibnamefont {Li}}, \bibinfo {author} {\bibfnamefont
  {L.}~\bibnamefont {Zeng}}, \bibinfo {author} {\bibfnamefont {D.-S.}\
  \bibnamefont {Ding}}, \bibinfo {author} {\bibfnamefont {B.-S.}\ \bibnamefont
  {Shi}}, \bibinfo {author} {\bibfnamefont {G.-C.}\ \bibnamefont {Guo}},\ and\
  \bibinfo {author} {\bibfnamefont {F.}~\bibnamefont {Nori}},\ }\bibfield
  {title} {\bibinfo {title} {All-optical reversible single-photon isolation at
  room temperature},\ }\href {https://doi.org/10.1126/sciadv.abe8924}
  {\bibfield  {journal} {\bibinfo  {journal} {Science Advances}\ }\textbf
  {\bibinfo {volume} {7}},\ \bibinfo {pages} {eabe8924} (\bibinfo {year}
  {2021})}\BibitemShut {NoStop}%
\bibitem [{\citenamefont {Aumiler}\ \emph {et~al.}(2005)\citenamefont
  {Aumiler}, \citenamefont {Ban}, \citenamefont
  {Skenderovi\ifmmode~\acute{c}\else \'{c}\fi{}},\ and\ \citenamefont
  {Pichler}}]{aumiler2005}%
  \BibitemOpen
  \bibfield  {author} {\bibinfo {author} {\bibfnamefont {D.}~\bibnamefont
  {Aumiler}}, \bibinfo {author} {\bibfnamefont {T.}~\bibnamefont {Ban}},
  \bibinfo {author} {\bibfnamefont {H.}~\bibnamefont
  {Skenderovi\ifmmode~\acute{c}\else \'{c}\fi{}}},\ and\ \bibinfo {author}
  {\bibfnamefont {G.}~\bibnamefont {Pichler}},\ }\bibfield  {title} {\bibinfo
  {title} {Velocity selective optical pumping of rb hyperfine lines induced by
  a train of femtosecond pulses},\ }\href
  {https://doi.org/10.1103/PhysRevLett.95.233001} {\bibfield  {journal}
  {\bibinfo  {journal} {Physical Review Letters}\ }\textbf {\bibinfo {volume}
  {95}},\ \bibinfo {pages} {233001} (\bibinfo {year} {2005})}\BibitemShut
  {NoStop}%
\bibitem [{\citenamefont {Afzelius}\ \emph {et~al.}(2009)\citenamefont
  {Afzelius}, \citenamefont {Simon}, \citenamefont {De~Riedmatten},\ and\
  \citenamefont {Gisin}}]{afzelius2009}%
  \BibitemOpen
  \bibfield  {author} {\bibinfo {author} {\bibfnamefont {M.}~\bibnamefont
  {Afzelius}}, \bibinfo {author} {\bibfnamefont {C.}~\bibnamefont {Simon}},
  \bibinfo {author} {\bibfnamefont {H.}~\bibnamefont {De~Riedmatten}},\ and\
  \bibinfo {author} {\bibfnamefont {N.}~\bibnamefont {Gisin}},\ }\bibfield
  {title} {\bibinfo {title} {Multimode quantum memory based on atomic frequency
  combs},\ }\href {https://doi.org/10.1103/PhysRevA.79.052329} {\bibfield
  {journal} {\bibinfo  {journal} {Physical Review A}\ }\textbf {\bibinfo
  {volume} {79}},\ \bibinfo {pages} {052329} (\bibinfo {year}
  {2009})}\BibitemShut {NoStop}%
\bibitem [{\citenamefont {Main}\ \emph {et~al.}(2021)\citenamefont {Main},
  \citenamefont {Hird}, \citenamefont {Gao}, \citenamefont {Walmsley},\ and\
  \citenamefont {Ledingham}}]{main2021}%
  \BibitemOpen
  \bibfield  {author} {\bibinfo {author} {\bibfnamefont {D.}~\bibnamefont
  {Main}}, \bibinfo {author} {\bibfnamefont {T.~M.}\ \bibnamefont {Hird}},
  \bibinfo {author} {\bibfnamefont {S.}~\bibnamefont {Gao}}, \bibinfo {author}
  {\bibfnamefont {I.~A.}\ \bibnamefont {Walmsley}},\ and\ \bibinfo {author}
  {\bibfnamefont {P.~M.}\ \bibnamefont {Ledingham}},\ }\bibfield  {title}
  {\bibinfo {title} {Room temperature atomic frequency comb storage for
  light},\ }\href {https://doi.org/10.1364/OL.426753} {\bibfield  {journal}
  {\bibinfo  {journal} {Optics Letters}\ }\textbf {\bibinfo {volume} {46}},\
  \bibinfo {pages} {2960} (\bibinfo {year} {2021})}\BibitemShut {NoStop}%
\bibitem [{\citenamefont {Sedlacek}\ \emph {et~al.}(2012)\citenamefont
  {Sedlacek}, \citenamefont {Schwettmann}, \citenamefont {K{\"u}bler},
  \citenamefont {L{\"o}w}, \citenamefont {Pfau},\ and\ \citenamefont
  {Shaffer}}]{sedlacek2012}%
  \BibitemOpen
  \bibfield  {author} {\bibinfo {author} {\bibfnamefont {J.~A.}\ \bibnamefont
  {Sedlacek}}, \bibinfo {author} {\bibfnamefont {A.}~\bibnamefont
  {Schwettmann}}, \bibinfo {author} {\bibfnamefont {H.}~\bibnamefont
  {K{\"u}bler}}, \bibinfo {author} {\bibfnamefont {R.}~\bibnamefont {L{\"o}w}},
  \bibinfo {author} {\bibfnamefont {T.}~\bibnamefont {Pfau}},\ and\ \bibinfo
  {author} {\bibfnamefont {J.~P.}\ \bibnamefont {Shaffer}},\ }\bibfield
  {title} {\bibinfo {title} {Microwave electrometry with {{Rydberg}} atoms in a
  vapour cell using bright atomic resonances},\ }\href
  {https://doi.org/10.1038/nphys2423} {\bibfield  {journal} {\bibinfo
  {journal} {Nature Physics}\ }\textbf {\bibinfo {volume} {8}},\ \bibinfo
  {pages} {819} (\bibinfo {year} {2012})}\BibitemShut {NoStop}%
\bibitem [{\citenamefont {Jing}\ \emph {et~al.}(2020)\citenamefont {Jing},
  \citenamefont {Hu}, \citenamefont {Ma}, \citenamefont {Zhang}, \citenamefont
  {Zhang}, \citenamefont {Xiao},\ and\ \citenamefont {Jia}}]{Jing2020}%
  \BibitemOpen
  \bibfield  {author} {\bibinfo {author} {\bibfnamefont {M.}~\bibnamefont
  {Jing}}, \bibinfo {author} {\bibfnamefont {Y.}~\bibnamefont {Hu}}, \bibinfo
  {author} {\bibfnamefont {J.}~\bibnamefont {Ma}}, \bibinfo {author}
  {\bibfnamefont {H.}~\bibnamefont {Zhang}}, \bibinfo {author} {\bibfnamefont
  {L.}~\bibnamefont {Zhang}}, \bibinfo {author} {\bibfnamefont
  {L.}~\bibnamefont {Xiao}},\ and\ \bibinfo {author} {\bibfnamefont
  {S.}~\bibnamefont {Jia}},\ }\bibfield  {title} {\bibinfo {title} {{Atomic
  superheterodyne receiver based on microwave-dressed Rydberg spectroscopy}},\
  }\href {https://doi.org/10.1038/s41567-020-0918-5} {\bibfield  {journal}
  {\bibinfo  {journal} {Nature Physics}\ }\textbf {\bibinfo {volume} {16}},\
  \bibinfo {pages} {911} (\bibinfo {year} {2020})}\BibitemShut {NoStop}%
\bibitem [{\citenamefont {Simons}\ \emph {et~al.}(2021)\citenamefont {Simons},
  \citenamefont {{Artusio-Glimpse}}, \citenamefont {Holloway}, \citenamefont
  {Imhof}, \citenamefont {Jefferts}, \citenamefont {Wyllie}, \citenamefont
  {Sawyer},\ and\ \citenamefont {Walker}}]{simons2021}%
  \BibitemOpen
  \bibfield  {author} {\bibinfo {author} {\bibfnamefont {M.~T.}\ \bibnamefont
  {Simons}}, \bibinfo {author} {\bibfnamefont {A.~B.}\ \bibnamefont
  {{Artusio-Glimpse}}}, \bibinfo {author} {\bibfnamefont {C.~L.}\ \bibnamefont
  {Holloway}}, \bibinfo {author} {\bibfnamefont {E.}~\bibnamefont {Imhof}},
  \bibinfo {author} {\bibfnamefont {S.~R.}\ \bibnamefont {Jefferts}}, \bibinfo
  {author} {\bibfnamefont {R.}~\bibnamefont {Wyllie}}, \bibinfo {author}
  {\bibfnamefont {B.~C.}\ \bibnamefont {Sawyer}},\ and\ \bibinfo {author}
  {\bibfnamefont {T.~G.}\ \bibnamefont {Walker}},\ }\bibfield  {title}
  {\bibinfo {title} {Continuous radio-frequency electric-field detection
  through adjacent rydberg resonance tuning},\ }\href
  {https://doi.org/10.1103/PhysRevA.104.032824} {\bibfield  {journal} {\bibinfo
   {journal} {Physical Review A}\ }\textbf {\bibinfo {volume} {104}},\ \bibinfo
  {pages} {3} (\bibinfo {year} {2021})}\BibitemShut {NoStop}%
\bibitem [{\citenamefont {Schlossberger}\ \emph {et~al.}(2024)\citenamefont
  {Schlossberger}, \citenamefont {Rotunno}, \citenamefont {{Artusio-Glimpse}},
  \citenamefont {Prajapati}, \citenamefont {Berweger}, \citenamefont {Shylla},
  \citenamefont {Simons},\ and\ \citenamefont {Holloway}}]{schlossberger2024}%
  \BibitemOpen
  \bibfield  {author} {\bibinfo {author} {\bibfnamefont {N.}~\bibnamefont
  {Schlossberger}}, \bibinfo {author} {\bibfnamefont {A.~P.}\ \bibnamefont
  {Rotunno}}, \bibinfo {author} {\bibfnamefont {A.}~\bibnamefont
  {{Artusio-Glimpse}}}, \bibinfo {author} {\bibfnamefont {N.}~\bibnamefont
  {Prajapati}}, \bibinfo {author} {\bibfnamefont {S.}~\bibnamefont {Berweger}},
  \bibinfo {author} {\bibfnamefont {D.}~\bibnamefont {Shylla}}, \bibinfo
  {author} {\bibfnamefont {M.~T.}\ \bibnamefont {Simons}},\ and\ \bibinfo
  {author} {\bibfnamefont {C.~L.}\ \bibnamefont {Holloway}},\ }\bibfield
  {title} {\bibinfo {title} {Zeeman-resolved autler-townes splitting in rydberg
  atoms with tunable resonances and a single transition dipole moment},\ }\href
  {https://doi.org/10.1103/PhysRevA.109.L021702} {\bibfield  {journal}
  {\bibinfo  {journal} {Physical Review A}\ }\textbf {\bibinfo {volume}
  {109}},\ \bibinfo {pages} {L021702} (\bibinfo {year} {2024})}\BibitemShut
  {NoStop}%
\bibitem [{\citenamefont {Elgee}\ \emph {et~al.}(2024)\citenamefont {Elgee},
  \citenamefont {Cox}, \citenamefont {Hill}, \citenamefont {Kunz},\ and\
  \citenamefont {Meyer}}]{elgee2024}%
  \BibitemOpen
  \bibfield  {author} {\bibinfo {author} {\bibfnamefont {P.~K.}\ \bibnamefont
  {Elgee}}, \bibinfo {author} {\bibfnamefont {K.~C.}\ \bibnamefont {Cox}},
  \bibinfo {author} {\bibfnamefont {J.~C.}\ \bibnamefont {Hill}}, \bibinfo
  {author} {\bibfnamefont {P.~D.}\ \bibnamefont {Kunz}},\ and\ \bibinfo
  {author} {\bibfnamefont {D.~H.}\ \bibnamefont {Meyer}},\ }\bibfield  {title}
  {\bibinfo {title} {Complete three-dimensional vector polarimetry with a
  rydberg-atom rf electrometer},\ }\href
  {https://doi.org/10.1103/PhysRevApplied.22.064012} {\bibfield  {journal}
  {\bibinfo  {journal} {Physical Review Applied}\ }\textbf {\bibinfo {volume}
  {22}},\ \bibinfo {pages} {064012} (\bibinfo {year} {2024})}\BibitemShut
  {NoStop}%
\bibitem [{\citenamefont {Robinson}\ \emph {et~al.}(2021)\citenamefont
  {Robinson}, \citenamefont {Prajapati}, \citenamefont {Senic}, \citenamefont
  {Simons},\ and\ \citenamefont {Holloway}}]{robinson2021determining}%
  \BibitemOpen
  \bibfield  {author} {\bibinfo {author} {\bibfnamefont {A.~K.}\ \bibnamefont
  {Robinson}}, \bibinfo {author} {\bibfnamefont {N.}~\bibnamefont {Prajapati}},
  \bibinfo {author} {\bibfnamefont {D.}~\bibnamefont {Senic}}, \bibinfo
  {author} {\bibfnamefont {M.~T.}\ \bibnamefont {Simons}},\ and\ \bibinfo
  {author} {\bibfnamefont {C.~L.}\ \bibnamefont {Holloway}},\ }\bibfield
  {title} {\bibinfo {title} {Determining the angle-of-arrival of a
  radio-frequency source with a rydberg atom-based sensor},\ }\href
  {https://doi.org/10.1063/5.0045601} {\bibfield  {journal} {\bibinfo
  {journal} {Applied Physics Letters}\ }\textbf {\bibinfo {volume} {118}},\
  \bibinfo {pages} {114001} (\bibinfo {year} {2021})}\BibitemShut {NoStop}%
\bibitem [{\citenamefont {Holloway}\ \emph {et~al.}(2014)\citenamefont
  {Holloway}, \citenamefont {Gordon}, \citenamefont {Schwarzkopf},
  \citenamefont {Anderson}, \citenamefont {Miller}, \citenamefont
  {Thaicharoen},\ and\ \citenamefont {Raithel}}]{holloway2014subwave}%
  \BibitemOpen
  \bibfield  {author} {\bibinfo {author} {\bibfnamefont {C.~L.}\ \bibnamefont
  {Holloway}}, \bibinfo {author} {\bibfnamefont {J.~A.}\ \bibnamefont
  {Gordon}}, \bibinfo {author} {\bibfnamefont {A.}~\bibnamefont {Schwarzkopf}},
  \bibinfo {author} {\bibfnamefont {D.~A.}\ \bibnamefont {Anderson}}, \bibinfo
  {author} {\bibfnamefont {S.~A.}\ \bibnamefont {Miller}}, \bibinfo {author}
  {\bibfnamefont {N.}~\bibnamefont {Thaicharoen}},\ and\ \bibinfo {author}
  {\bibfnamefont {G.}~\bibnamefont {Raithel}},\ }\bibfield  {title} {\bibinfo
  {title} {Sub-wavelength imaging and field mapping via electromagnetically
  induced transparency and autler-townes splitting in rydberg atoms},\ }\href
  {https://doi.org/10.1063/1.4883635} {\bibfield  {journal} {\bibinfo
  {journal} {Applied Physics Letters}\ }\textbf {\bibinfo {volume} {104}},\
  \bibinfo {pages} {244102} (\bibinfo {year} {2014})}\BibitemShut {NoStop}%
\bibitem [{\citenamefont {Meyer}\ \emph {et~al.}(2021)\citenamefont {Meyer},
  \citenamefont {O'Brien}, \citenamefont {Fahey}, \citenamefont {Cox},\ and\
  \citenamefont {Kunz}}]{meyer2021}%
  \BibitemOpen
  \bibfield  {author} {\bibinfo {author} {\bibfnamefont {D.~H.}\ \bibnamefont
  {Meyer}}, \bibinfo {author} {\bibfnamefont {C.}~\bibnamefont {O'Brien}},
  \bibinfo {author} {\bibfnamefont {D.~P.}\ \bibnamefont {Fahey}}, \bibinfo
  {author} {\bibfnamefont {K.~C.}\ \bibnamefont {Cox}},\ and\ \bibinfo {author}
  {\bibfnamefont {P.~D.}\ \bibnamefont {Kunz}},\ }\bibfield  {title} {\bibinfo
  {title} {Optimal atomic quantum sensing using
  electromagnetically-induced-transparency readout},\ }\href
  {https://doi.org/10.1103/PhysRevA.104.043103} {\bibfield  {journal} {\bibinfo
   {journal} {Physical Review A}\ }\textbf {\bibinfo {volume} {104}},\ \bibinfo
  {pages} {043103} (\bibinfo {year} {2021})}\BibitemShut {NoStop}%
\bibitem [{\citenamefont {Cox}\ \emph {et~al.}(2018)\citenamefont {Cox},
  \citenamefont {Meyer}, \citenamefont {Fatemi},\ and\ \citenamefont
  {Kunz}}]{cox2018Quantum}%
  \BibitemOpen
  \bibfield  {author} {\bibinfo {author} {\bibfnamefont {K.~C.}\ \bibnamefont
  {Cox}}, \bibinfo {author} {\bibfnamefont {D.~H.}\ \bibnamefont {Meyer}},
  \bibinfo {author} {\bibfnamefont {F.~K.}\ \bibnamefont {Fatemi}},\ and\
  \bibinfo {author} {\bibfnamefont {P.~D.}\ \bibnamefont {Kunz}},\ }\bibfield
  {title} {\bibinfo {title} {Quantum-limited atomic receiver in the
  electrically small regime},\ }\href
  {https://doi.org/10.1103/PhysRevLett.121.110502} {\bibfield  {journal}
  {\bibinfo  {journal} {Physical Review Letters}\ }\textbf {\bibinfo {volume}
  {121}},\ \bibinfo {pages} {110502} (\bibinfo {year} {2018})}\BibitemShut
  {NoStop}%
\bibitem [{\citenamefont {Anderson}\ \emph {et~al.}(2021)\citenamefont
  {Anderson}, \citenamefont {Sapiro},\ and\ \citenamefont
  {Raithel}}]{anderson2021}%
  \BibitemOpen
  \bibfield  {author} {\bibinfo {author} {\bibfnamefont {D.~A.}\ \bibnamefont
  {Anderson}}, \bibinfo {author} {\bibfnamefont {R.~E.}\ \bibnamefont
  {Sapiro}},\ and\ \bibinfo {author} {\bibfnamefont {G.}~\bibnamefont
  {Raithel}},\ }\bibfield  {title} {\bibinfo {title} {An atomic receiver for am
  and fm radio communication},\ }\href
  {https://doi.org/10.1109/TAP.2020.2987112} {\bibfield  {journal} {\bibinfo
  {journal} {IEEE Transactions on Antennas and Propagation}\ }\textbf {\bibinfo
  {volume} {69}},\ \bibinfo {pages} {2455} (\bibinfo {year}
  {2021})}\BibitemShut {NoStop}%
\bibitem [{\citenamefont {Simons}\ \emph {et~al.}(2019)\citenamefont {Simons},
  \citenamefont {Haddab}, \citenamefont {Gordon},\ and\ \citenamefont
  {Holloway}}]{simons2019b}%
  \BibitemOpen
  \bibfield  {author} {\bibinfo {author} {\bibfnamefont {M.~T.}\ \bibnamefont
  {Simons}}, \bibinfo {author} {\bibfnamefont {A.~H.}\ \bibnamefont {Haddab}},
  \bibinfo {author} {\bibfnamefont {J.~A.}\ \bibnamefont {Gordon}},\ and\
  \bibinfo {author} {\bibfnamefont {C.~L.}\ \bibnamefont {Holloway}},\
  }\bibfield  {title} {\bibinfo {title} {A rydberg atom-based mixer: Measuring
  the phase of a radio frequency wave},\ }\href
  {https://doi.org/10.1063/1.5088821} {\bibfield  {journal} {\bibinfo
  {journal} {Applied Physics Letters}\ }\textbf {\bibinfo {volume} {114}},\
  \bibinfo {pages} {114101} (\bibinfo {year} {2019})}\BibitemShut {NoStop}%
\bibitem [{\citenamefont {Holloway}\ \emph {et~al.}(2019)\citenamefont
  {Holloway}, \citenamefont {Simons}, \citenamefont {Gordon},\ and\
  \citenamefont {Novotny}}]{holloway2019}%
  \BibitemOpen
  \bibfield  {author} {\bibinfo {author} {\bibfnamefont {C.~L.}\ \bibnamefont
  {Holloway}}, \bibinfo {author} {\bibfnamefont {M.~T.}\ \bibnamefont
  {Simons}}, \bibinfo {author} {\bibfnamefont {J.~A.}\ \bibnamefont {Gordon}},\
  and\ \bibinfo {author} {\bibfnamefont {D.}~\bibnamefont {Novotny}},\
  }\bibfield  {title} {\bibinfo {title} {Detecting and receiving
  phase-modulated signals with a rydberg atom-based receiver},\ }\href
  {https://doi.org/10.1109/LAWP.2019.2931450} {\bibfield  {journal} {\bibinfo
  {journal} {IEEE Antennas and Wireless Propagation Letters}\ }\textbf
  {\bibinfo {volume} {18}},\ \bibinfo {pages} {1853} (\bibinfo {year}
  {2019})}\BibitemShut {NoStop}%
\bibitem [{\citenamefont {Carr}\ \emph {et~al.}(2013)\citenamefont {Carr},
  \citenamefont {Ritter}, \citenamefont {Wade}, \citenamefont {Adams},\ and\
  \citenamefont {Weatherill}}]{Carr2013}%
  \BibitemOpen
  \bibfield  {author} {\bibinfo {author} {\bibfnamefont {C.}~\bibnamefont
  {Carr}}, \bibinfo {author} {\bibfnamefont {R.}~\bibnamefont {Ritter}},
  \bibinfo {author} {\bibfnamefont {C.~G.}\ \bibnamefont {Wade}}, \bibinfo
  {author} {\bibfnamefont {C.~S.}\ \bibnamefont {Adams}},\ and\ \bibinfo
  {author} {\bibfnamefont {K.~J.}\ \bibnamefont {Weatherill}},\ }\bibfield
  {title} {\bibinfo {title} {Nonequilibrium phase transition in a dilute
  rydberg ensemble},\ }\href {https://doi.org/10.1103/PhysRevLett.111.113901}
  {\bibfield  {journal} {\bibinfo  {journal} {Physical Review Letters}\
  }\textbf {\bibinfo {volume} {111}},\ \bibinfo {pages} {113901} (\bibinfo
  {year} {2013})}\BibitemShut {NoStop}%
\bibitem [{\citenamefont {Yuan}\ \emph {et~al.}(2023)\citenamefont {Yuan},
  \citenamefont {Yang}, \citenamefont {Jing}, \citenamefont {Zhang},
  \citenamefont {Jiao}, \citenamefont {Li}, \citenamefont {Zhang},
  \citenamefont {Xiao},\ and\ \citenamefont
  {Jia}}]{yuanQuantumSensingMicrowave2023}%
  \BibitemOpen
  \bibfield  {author} {\bibinfo {author} {\bibfnamefont {J.}~\bibnamefont
  {Yuan}}, \bibinfo {author} {\bibfnamefont {W.}~\bibnamefont {Yang}}, \bibinfo
  {author} {\bibfnamefont {M.}~\bibnamefont {Jing}}, \bibinfo {author}
  {\bibfnamefont {H.}~\bibnamefont {Zhang}}, \bibinfo {author} {\bibfnamefont
  {Y.}~\bibnamefont {Jiao}}, \bibinfo {author} {\bibfnamefont {W.}~\bibnamefont
  {Li}}, \bibinfo {author} {\bibfnamefont {L.}~\bibnamefont {Zhang}}, \bibinfo
  {author} {\bibfnamefont {L.}~\bibnamefont {Xiao}},\ and\ \bibinfo {author}
  {\bibfnamefont {S.}~\bibnamefont {Jia}},\ }\bibfield  {title} {\bibinfo
  {title} {Quantum sensing of microwave electric fields based on {{Rydberg}}
  atoms},\ }\href {https://doi.org/10.1088/1361-6633/acf22f} {\bibfield
  {journal} {\bibinfo  {journal} {Reports On Progress In Physics}\ }\textbf
  {\bibinfo {volume} {86}},\ \bibinfo {pages} {106001} (\bibinfo {year}
  {2023})}\BibitemShut {NoStop}%
\bibitem [{\citenamefont {Schlossberger}(2024)}]{schlossberger2024a}%
  \BibitemOpen
  \bibfield  {author} {\bibinfo {author} {\bibfnamefont {N.}~\bibnamefont
  {Schlossberger}},\ }\bibfield  {title} {\bibinfo {title} {Rydberg states of
  alkali atoms in atomic vapour as {{SI-traceable}} field probes and
  communications receivers},\ }\href
  {https://doi.org/10.1038/s42254-024-00756-7} {\bibfield  {journal} {\bibinfo
  {journal} {Nature Reviews Physics}\ }\textbf {\bibinfo {volume} {6}},\
  \bibinfo {pages} {606} (\bibinfo {year} {2024})}\BibitemShut {NoStop}%
\bibitem [{\citenamefont {Prajapati}\ \emph {et~al.}(2021)\citenamefont
  {Prajapati}, \citenamefont {Robinson}, \citenamefont {Berweger},
  \citenamefont {Simons}, \citenamefont {Artusio-Glimpse},\ and\ \citenamefont
  {Holloway}}]{prajapati2021}%
  \BibitemOpen
  \bibfield  {author} {\bibinfo {author} {\bibfnamefont {N.}~\bibnamefont
  {Prajapati}}, \bibinfo {author} {\bibfnamefont {A.~K.}\ \bibnamefont
  {Robinson}}, \bibinfo {author} {\bibfnamefont {S.}~\bibnamefont {Berweger}},
  \bibinfo {author} {\bibfnamefont {M.~T.}\ \bibnamefont {Simons}}, \bibinfo
  {author} {\bibfnamefont {A.~B.}\ \bibnamefont {Artusio-Glimpse}},\ and\
  \bibinfo {author} {\bibfnamefont {C.~L.}\ \bibnamefont {Holloway}},\
  }\bibfield  {title} {\bibinfo {title} {Enhancement of electromagnetically
  induced transparency based rydberg-atom electrometry through population
  repumping},\ }\href {https://doi.org/10.1063/5.0069195} {\bibfield  {journal}
  {\bibinfo  {journal} {Applied Physics Letters}\ }\textbf {\bibinfo {volume}
  {119}},\ \bibinfo {pages} {214001} (\bibinfo {year} {2021})}\BibitemShut
  {NoStop}%
\bibitem [{\citenamefont {Cai}\ \emph {et~al.}(2023)\citenamefont {Cai},
  \citenamefont {You}, \citenamefont {Zhang}, \citenamefont {Xu},\ and\
  \citenamefont {Liu}}]{cai2023a}%
  \BibitemOpen
  \bibfield  {author} {\bibinfo {author} {\bibfnamefont {M.}~\bibnamefont
  {Cai}}, \bibinfo {author} {\bibfnamefont {S.}~\bibnamefont {You}}, \bibinfo
  {author} {\bibfnamefont {S.}~\bibnamefont {Zhang}}, \bibinfo {author}
  {\bibfnamefont {Z.}~\bibnamefont {Xu}},\ and\ \bibinfo {author}
  {\bibfnamefont {H.}~\bibnamefont {Liu}},\ }\bibfield  {title} {\bibinfo
  {title} {Sensitivity extension of atom-based amplitude-modulation microwave
  electrometry via high rydberg states},\ }\href
  {https://doi.org/10.1063/5.0146768} {\bibfield  {journal} {\bibinfo
  {journal} {Applied Physics Letters}\ }\textbf {\bibinfo {volume} {122}},\
  \bibinfo {pages} {161103} (\bibinfo {year} {2023})}\BibitemShut {NoStop}%
\bibitem [{\citenamefont {Wu}\ \emph {et~al.}(2025{\natexlab{a}})\citenamefont
  {Wu}, \citenamefont {Mao}, \citenamefont {Sang}, \citenamefont {Sun},
  \citenamefont {Liu}, \citenamefont {Lin}, \citenamefont {An},\ and\
  \citenamefont {Fu}}]{wu2025}%
  \BibitemOpen
  \bibfield  {author} {\bibinfo {author} {\bibfnamefont {B.}~\bibnamefont
  {Wu}}, \bibinfo {author} {\bibfnamefont {R.}~\bibnamefont {Mao}}, \bibinfo
  {author} {\bibfnamefont {D.}~\bibnamefont {Sang}}, \bibinfo {author}
  {\bibfnamefont {Z.}~\bibnamefont {Sun}}, \bibinfo {author} {\bibfnamefont
  {Y.}~\bibnamefont {Liu}}, \bibinfo {author} {\bibfnamefont {Y.}~\bibnamefont
  {Lin}}, \bibinfo {author} {\bibfnamefont {Q.}~\bibnamefont {An}},\ and\
  \bibinfo {author} {\bibfnamefont {Y.}~\bibnamefont {Fu}},\ }\bibfield
  {title} {\bibinfo {title} {Enhancing sensitivity of atomic microwave
  receivers based on optimal laser arrays},\ }\href
  {https://doi.org/10.1109/TAP.2024.3486553} {\bibfield  {journal} {\bibinfo
  {journal} {IEEE Transactions on Antennas and Propagation}\ }\textbf {\bibinfo
  {volume} {73}},\ \bibinfo {pages} {793} (\bibinfo {year}
  {2025}{\natexlab{a}})}\BibitemShut {NoStop}%
\bibitem [{\citenamefont {Wu}\ \emph {et~al.}(2025{\natexlab{b}})\citenamefont
  {Wu}, \citenamefont {Liao}, \citenamefont {Sang}, \citenamefont {Liu},\ and\
  \citenamefont {Fu}}]{wu2025a}%
  \BibitemOpen
  \bibfield  {author} {\bibinfo {author} {\bibfnamefont {B.}~\bibnamefont
  {Wu}}, \bibinfo {author} {\bibfnamefont {D.}~\bibnamefont {Liao}}, \bibinfo
  {author} {\bibfnamefont {D.}~\bibnamefont {Sang}}, \bibinfo {author}
  {\bibfnamefont {Y.}~\bibnamefont {Liu}},\ and\ \bibinfo {author}
  {\bibfnamefont {Y.}~\bibnamefont {Fu}},\ }\bibfield  {title} {\bibinfo
  {title} {Enhancing sensitivity of an atomic microwave receiver via a
  fabry-perot cavity},\ }\href {https://doi.org/10.1109/TAP.2024.3480459}
  {\bibfield  {journal} {\bibinfo  {journal} {IEEE Transactions on Antennas and
  Propagation}\ }\textbf {\bibinfo {volume} {73}},\ \bibinfo {pages} {863}
  (\bibinfo {year} {2025}{\natexlab{b}})}\BibitemShut {NoStop}%
\bibitem [{\citenamefont {Yan}\ and\ \citenamefont {Hase}(2000)}]{yan2000}%
  \BibitemOpen
  \bibfield  {author} {\bibinfo {author} {\bibfnamefont {T.}~\bibnamefont
  {Yan}}\ and\ \bibinfo {author} {\bibfnamefont {W.~L.}\ \bibnamefont {Hase}},\
  }\bibfield  {title} {\bibinfo {title} {Origin of the boltzmann translational
  energy distribution in the scattering of hyperthermal ne atoms off a
  self-assembled monolayer},\ }\href {https://doi.org/10.1039/a908370g}
  {\bibfield  {journal} {\bibinfo  {journal} {Physical Chemistry Chemical
  Physics}\ }\textbf {\bibinfo {volume} {2}},\ \bibinfo {pages} {901} (\bibinfo
  {year} {2000})}\BibitemShut {NoStop}%
\bibitem [{\citenamefont {Chen}\ \emph {et~al.}(2020)\citenamefont {Chen},
  \citenamefont {Lim}, \citenamefont {Huang}, \citenamefont {Dumke},\ and\
  \citenamefont {Lan}}]{chen2020}%
  \BibitemOpen
  \bibfield  {author} {\bibinfo {author} {\bibfnamefont {Z.}~\bibnamefont
  {Chen}}, \bibinfo {author} {\bibfnamefont {H.~M.}\ \bibnamefont {Lim}},
  \bibinfo {author} {\bibfnamefont {C.}~\bibnamefont {Huang}}, \bibinfo
  {author} {\bibfnamefont {R.}~\bibnamefont {Dumke}},\ and\ \bibinfo {author}
  {\bibfnamefont {S.-Y.}\ \bibnamefont {Lan}},\ }\bibfield  {title} {\bibinfo
  {title} {Quantum-enhanced velocimetry with doppler-broadened atomic vapor},\
  }\href {https://doi.org/10.1103/PhysRevLett.124.093202} {\bibfield  {journal}
  {\bibinfo  {journal} {Physical Review Letters}\ }\textbf {\bibinfo {volume}
  {124}},\ \bibinfo {pages} {093202} (\bibinfo {year} {2020})}\BibitemShut
  {NoStop}%
\bibitem [{\citenamefont {Wade}\ \emph {et~al.}(2017)\citenamefont {Wade},
  \citenamefont {{\v S}ibali{\'c}}, \citenamefont {{de Melo}}, \citenamefont
  {Kondo}, \citenamefont {Adams},\ and\ \citenamefont
  {Weatherill}}]{wade2017a}%
  \BibitemOpen
  \bibfield  {author} {\bibinfo {author} {\bibfnamefont {C.~G.}\ \bibnamefont
  {Wade}}, \bibinfo {author} {\bibfnamefont {N.}~\bibnamefont {{\v
  S}ibali{\'c}}}, \bibinfo {author} {\bibfnamefont {N.~R.}\ \bibnamefont {{de
  Melo}}}, \bibinfo {author} {\bibfnamefont {J.~M.}\ \bibnamefont {Kondo}},
  \bibinfo {author} {\bibfnamefont {C.~S.}\ \bibnamefont {Adams}},\ and\
  \bibinfo {author} {\bibfnamefont {K.~J.}\ \bibnamefont {Weatherill}},\
  }\bibfield  {title} {\bibinfo {title} {Real-time near-field terahertz imaging
  with atomic optical fluorescence},\ }\href
  {https://doi.org/10.1038/nphoton.2016.214} {\bibfield  {journal} {\bibinfo
  {journal} {Nature Photonics}\ }\textbf {\bibinfo {volume} {11}},\ \bibinfo
  {pages} {40} (\bibinfo {year} {2017})}\BibitemShut {NoStop}%
\bibitem [{\citenamefont {Downes}\ \emph {et~al.}(2020)\citenamefont {Downes},
  \citenamefont {MacKellar}, \citenamefont {Whiting}, \citenamefont
  {Bourgenot}, \citenamefont {Adams},\ and\ \citenamefont
  {Weatherill}}]{downes2020b}%
  \BibitemOpen
  \bibfield  {author} {\bibinfo {author} {\bibfnamefont {L.~A.}\ \bibnamefont
  {Downes}}, \bibinfo {author} {\bibfnamefont {A.~R.}\ \bibnamefont
  {MacKellar}}, \bibinfo {author} {\bibfnamefont {D.~J.}\ \bibnamefont
  {Whiting}}, \bibinfo {author} {\bibfnamefont {C.}~\bibnamefont {Bourgenot}},
  \bibinfo {author} {\bibfnamefont {C.~S.}\ \bibnamefont {Adams}},\ and\
  \bibinfo {author} {\bibfnamefont {K.~J.}\ \bibnamefont {Weatherill}},\
  }\bibfield  {title} {\bibinfo {title} {Full-{{Field Terahertz Imaging}} at
  {{Kilohertz Frame Rates Using Atomic Vapor}}},\ }\href
  {https://doi.org/10.1103/PhysRevX.10.011027} {\bibfield  {journal} {\bibinfo
  {journal} {Physical Review X}\ }\textbf {\bibinfo {volume} {10}},\ \bibinfo
  {pages} {011027} (\bibinfo {year} {2020})}\BibitemShut {NoStop}%
\end{thebibliography}%
	
\end{document}